\documentstyle[aps,pre,eqsecnum]{revtex}

\draft

\begin{document}

\title{Quantum Kramers' equation for energy diffusion and barrier crossing
dynamics in the low friction regime}

\author{Dhruba~Banerjee$^1$, 
Suman~Kumar~Banik$^1${\footnote{
Present Address: Max-Planck-Institut f\"ur Physik komplexer Systeme,
N\"othnitzer Stra\ss e 38, 01187 Dresden, Germany}}, 
Bidhan~Chandra~Bag$^2$ 
and Deb~Shankar~Ray$^1${\footnote{Email address: pcdsr@mahendra.iacs.res.in}}
}

\address{$^1$Indian Association for the Cultivation of Science,
Jadavpur, Calcutta 700 032, India\\
$^2$Department of Chemistry, Visva-Bharati, Shantiniketan 731 235, India}

\date{Dated: August 7, 2002}

\maketitle

\begin{abstract}
Based on a true phase space probability distribution function and an ensemble
averaging procedure we have recently developed
[ Phys. Rev. E {\bf 65}, 021109 (2002) ] a non-Markovian quantum Kramers'
equation to derive the quantum rate coefficient 
for barrier crossing due to thermal activation and tunneling
in the intermediate to strong friction regime. 
We complement and extend this approach to weak
friction regime to derive quantum Kramers' equation in energy space and
the rate of decay from a metastable well. The theory is valid for arbitrary 
temperature and noise correlation. We show that depending on the nature
of the potential there may be a net reduction of the total quantum rate below
its corresponding classical value which is in conformity with earlier
observation. The method is independent of path integral approaches and
takes care of quantum effects to all orders.
\end{abstract}



\section{Introduction}

The dynamics of noise-induced rate processes was first successfully treated
in a seminal paper by Kramers in 1940 \cite{hak}. With the advances in
experimental methods for monitoring ultrafast processes on microscopic
spatial and temporal scales over the last two decades \cite{diau,castelman},
this has been the subject of numerous investigation from classical,
semiclassical and quantum mechanical point of view \cite{htb,vim,fh,th1}.
The classical Kramers' theory has thus been extended to non-Markovian 
dissipation models \cite{gh,hm,cn},
generalization to complex potential \cite{gt,stein,bcb}
and to many degrees of freedom \cite{jsl,bpz},
fluctuating barrier problem \cite{dg} 
and non-stationary activated processes \cite{nsp},
thermal ratchet \cite{tr,ken} and molecular motors \cite{mm},
analysis of quantum \cite{htb,vim,garg,ingold,uw} 
and semiclassical effects \cite{jrc}, 
calculation of time-dependent transmission coefficient 
\cite{kt,srl}, fractional kinetics \cite{mk,rsl},
nonequilibrium open systems \cite{skb,endb}, activationless escape of a
free Brownian particle \cite{bicout} and other related issues 
\cite{htb,fh,th1}.

Although classical Kramers' equation was proposed more than sixty years ago
and quantum Kramers' problem of escape from a metastable state has attracted
wide attention over the last two decades \cite{htb}, the quantum version
of Kramers' equation was not reported in the literature. This is probably
because of the fact that the traditional method of treatment of quantum
Kramers' problem rests on calculation of partition function for 
a system-reservoir Hamiltonian in terms of path integrals, rather than on 
evolution of probability distribution function as used in classical theory
of stochastic processes.
Very recently we
have developed \cite{dbbr} a method based on true quantum ($c$-number)
phase space distribution function (rather than quasi-probability function,
like Wigner function \cite{epw}) to derive for the first time an exact
non-Markovian quantum Kramers' equation which is valid for arbitrary
temperature and friction. The solution of this equation as an appropriate
boundary value problem results in an expression for quantum rate
coefficient which not only reduces to Kramers'-Grote-Hynes \cite{hak,gh}
rate in the classical limit but also to the result corresponding to
zero-temperature tunneling in the full quantum limit, treated by
Caldeira and Leggett \cite{aoc} in early eighties. 
The rate coefficient thus derived pertains
to spatially-diffusion limited processes and is therefore valid
for intermediate to strong friction regime. We undertake the present study
with the following specific objectives to complement this work in the low 
friction regime where the process is controlled by energy diffusion.

\noindent
(1) To extend the treatment of quantum Kramers' problem for low to 
low-moderate friction we develop a quantum Kramers' equation for energy 
diffusion which is a quantum version of classical non-Markovian
equation of Carmeli and Nitzan (CN) \cite{cn} proposed in early eighties.

\noindent
(2) Our aim here is the inclusion of memory effects for arbitrary noise
correlation of the heat bath kept at an arbitrary temperature taking into
consideration of the quantum effects (corrections) to all orders.

\noindent
(3) We solve the quantum Kramers' equation for energy diffusion to derive
an explicit form of rate coefficient in the weak friction regime and
show that it reduces to non-Markovian counterpart of H\"anggi and Weiss 
\cite{hw} in the classical limit. Furthermore it provides the rate
coefficient of low temperature tunneling (down to absolute zero) 
in the quantum limit. The present theory 
thus interpolates between thermal activation and tunneling for weak 
dissipation within a single scheme and is a direct extension of classical 
theory to quantum domain.

The classical non-Markovian Fokker-Planck equation in the energy variable
for arbitrary noise correlation was first proposed by CN \cite{cn}. 
The detailed classical analysis by several groups
\cite{cn,hw,gh1,th} revealed that the 
rate, in general, is significantly modified by memory effects when compared 
to corresponding Kramers' theory in the static friction limit. 
As mentioned earlier, the traditional quantum 
treatment of the Kramers' problem in weak friction limit on the other hand, 
is based on functional integral approach \cite{htb} which takes care of 
dissipative tunneling \cite{aoc}. Since for weak friction limit at a finite 
temperature one finds a small population at the upper energy levels of the 
system which results in non-equilibrium effects, quantum correction to 
classical Kramers' weak damping results above the cross-over temperature is 
of considerable interest. Several authors \cite{vm,lo,rj,chow,dekker,griff} 
have addressed this problem in relation to nonequilibrium quantum tunneling 
out of a metastable state. Although the method of functional integrals as 
employed in these theories has been successful in treating arbitrary coupling 
and correlation time scales in a formally exact manner, analytic evaluation
of these integrals usually requires semi-classical approximations, e.g.,
semiclassical steepest descent method, WKB approach, etc., or other specific 
cases, which put restriction on applicability 
of the theories in several ways.
Secondly, the weak coupling theories $($ which have been
extensively used
in quantum optics \cite{whl} since seventies $)$ based on master equations
that make use of quasi-probability functions, like Wigner function 
\cite{epw} often pose serious difficulties concerning negativity 
or singularity of the probability distribution functions
as discussed in detail in earlier work \cite{bbr}. Third, when the system
or the system-reservoir coupling is nonlinear, the differential equations
concern higher (than second) derivatives of quasi-probability functions
\cite{rv} for which quantum-classical correspondence gets blurred. 
Our approach here is based on {\it true quantum
probability phase space function} and is free from such difficulties.
Furthermore an important decisive advantage of the scheme is that it 
allows us to implement the classical non-Markov theories of activated
processes in a full quantum setting without taking recourse to any
semiclassical technique. In what follows we
specifically apply the classical procedure of Lax \cite{lax},
CN \cite{cn}, B\"uttiker, Harris and Landauer (BHL) \cite{bhl},
H\"anggi and Weiss \cite{hw} in our quantum phase space formulation to
develop a non-Markovian quantum Kramers' equation in energy variable
and derive an expression for quantum rate coefficient in the spirit
of classical Kramers' theory. The quantum Kramers' equation and the
rate coefficient are classical looking in form but quantum mechanical
in their content and it is easy to recover their non-Markovian classical
counterparts in the limit $\hbar \rightarrow 0$.

The Kramers' kinetics in the low friction regime is just not
a theoretical issue today but has been a subject of experimental
investigation over the last two decades \cite{htb,troe,hej,jt,hara,cmc}. 
A number of experimental work in chemistry aimed at detecting Kramers'
turnover phenomena, in various reactions which can be conveniently 
explained in terms
of a one-dimensional model, e.g., iodine atom recombination in various
inert solvents \cite{troe}, 
chair-chair isomerization of cyclohexane \cite{hej},
excited state isomerization of 2-alkylanthracene \cite{hara}. Another class
of experiments where the energy diffusion mechanism has been successfully
implemented concern resonant activation of a Josephson junction
\cite{th,htb} and decay of zero voltage state in a current-biased
Josephson junction \cite{cmc}.  
The non-exponential decay behaviour in spin glass \cite{htb} is also an
area of active research in this context.
All these problems 
have their quantum counterparts which are being considered for 
further studies in rate theory although the 
experimental evidence of some of the 
theoretical predictions in low temperature quantum effects in weak friction 
regime is still awaited.

The outlay of the paper is as follows. We introduce a $c$-number
representation of a generalized quantum Langevin equation in Sec.~{II}.
This formulation helps us to use the classical formulation of CN \cite{cn}
for deriving a non-Markovian quantum Kramers' equation in energy space in
Sec.~{III}. We solve the problem of quantum energy diffusion controlled
rate coefficient in the spirit of classical theory \cite{hw,bhl} in
Sec.~{IV}. This reduces to classical rate expression of
Kramers-H\"anggi-Weiss \cite{hw} form in the limit $\hbar \rightarrow 0$.
An explicit example with a cubic potential is worked out to illustrate
the theory in Sec.~{V}. The paper is concluded in Sec.~{VI}.


\section{The quantum Generalized Langevin equation (QGLE) in $c$-numbers }

We consider a particle in a medium. The latter is modeled as a set of
harmonic oscillators with frequency $\{ \omega_i \}$. Evolution of such a 
quantum open system has been studied over the last several decades under a 
variety of reasonable assumptions. Specifically our interest here is to
develop an exact 
description of quantum Brownian motion within the perview of this model
described by the following Hamiltonian \cite{rz},

\begin{equation}
\label{eqn1}
\hat{H} = \frac{ \hat{P}^2 }{2} + V ( \hat{X} ) 
+ \sum_j \left [   \frac{ \hat{p}_j^2 }{2}
+ \frac{1}{2} \kappa_j ( \hat{q}_j - \hat{X} )^2  \right ] \; \; .
\end{equation}

\noindent
Here $\hat{X}$ and $\hat{P}$ are co-ordinate and momentum operators of
the particle and the set $ \{ \hat{q}_j,\hat{p}_j \}$ is the set of 
co-ordinate and momentum operators for the reservoir oscillators coupled 
linearly to the system through their coupling coefficients $\kappa_j$. 
The potential $V(\hat{X})$ is due to the external force field for the 
Brownian particle. The co-ordinate and momentum operators follow the
usual commutation relation [$\hat{X}, \hat{P}$] = $i\hbar$ and
[$\hat{q}_j, \hat{p}_j$] = $i\hbar \delta_{ij}$.
Note that in writing down the Hamiltonian no rotating 
wave approximation has been used.

Eliminating the reservoir degrees of freedom in the usual way 
\cite{whl,west,jkb} we obtain the
operator Langevin equation for the particle,
\begin{equation}
\label{eqn2}
\ddot{ \hat{X} } (t) + \int_0^t dt'  \beta (t-t')  \dot{ \hat{X} } (t')
+ V' ( \hat{X} )
= \hat{F} (t) \; \; ,
\end{equation}

\noindent
where the noise operator $\hat{F} (t)$ and the memory kernel 
$\beta (t)$ are given by
\begin{equation}
\label{eqn3}
\hat{F} (t) = \sum_j \left [  
\left \{ \hat{q}_j (0) - \hat{X} (0) \right \} 
\kappa_j  \cos \omega_j t +
\hat{p}_j (0)  \kappa_j^{1/2}  \sin \omega_j t  \right ]
\end{equation}

\noindent
and
\begin{equation}
\label{eqn4}
\beta (t) = \sum_j \kappa_j  \cos \omega_j t \; \; ,
\end{equation}

\noindent
with $\kappa_j = \omega_j^2$ ( masses have been assumed to be unity ). 

The Eq.(\ref{eqn2}) is an exact quantized operator Langevin equation which is
now a standard textbook material \cite{whl} and for which
the noise properties
of $\hat{F} (t)$ can be defined using a suitable initial canonical 
distribution of the bath co-ordinates and momenta. Our aim here is 
to replace it by an equivalent QGLE in $c$-numbers. Again this is not a new 
problem so long as one is restricted to standard quasi-probabilistic methods 
using, for example, Wigner functions \cite{epw}. 
To address the problem of quantum non-Markovian dynamics in terms of a 
{\it true probabilistic description} 
we, however, follow a different procedure. 
We {\it first} carry out the 
{\it quantum mechanical average} of Eq.(\ref{eqn2})
\begin{equation}
\label{eqn5}
\langle \ddot{ \hat{X} } (t) \rangle + 
\int_0^t dt'  \beta (t-t')  \langle \dot{ \hat{X} } (t') \rangle
+ \langle V' ( \hat{X} ) \rangle
= \langle \hat{F} (t) \rangle
\end{equation}

\noindent
where the average $\langle \ldots \rangle$
is taken over the initial product separable quantum states
of the particle and the bath oscillators at $t=0$,
$| \phi \rangle \{ | \alpha_1 \rangle | \alpha_2 \rangle \ldots
| \alpha_N \rangle \} $.
Here $| \phi \rangle$ denotes any arbitrary 
initial state of the particle and
$| \alpha_i \rangle$ corresponds to the initial
coherent state of the $i$-th bath oscillator. $|\alpha_i \rangle$
is given by 
$|\alpha_i \rangle = \exp(-|\alpha_i|^2/2) 
\sum_{n_i=0}^\infty (\alpha_i^{n_i} /\sqrt{n_i !} ) | n_i \rangle $,
$\alpha_i$ being expressed in terms of the mean values of the co-ordinate
and momentum of the $i$-th oscillator,
$\langle \hat{q}_i (0) \rangle = ( \sqrt{\hbar} /2\omega_i)                                                                                
(\alpha_i + \alpha_i^\star )$ and
$\langle \hat{p}_i (0) \rangle = i \sqrt{\hbar\omega_i/2 }
(\alpha_i^\star - \alpha_i )$, respectively.
It is important to note that $\langle \hat{F} (t) \rangle$
of Eq.(\ref{eqn5}) is a classical-like noise term which, in general, is a
non-zero number because of the quantum mechanical averaging over the 
co-ordinate and momentum operators of the bath oscillators with respect to 
the initial coherent states and arbitrary initial state of the particle
and is given by
\begin{equation}
\label{eqn6}
\langle \hat{F} (t) \rangle = \sum_j \left [  
\left \{  \langle \hat{q}_j (0) \rangle - \langle \hat{X} (0) \rangle 
\right \}  \kappa_j  \cos \omega_j t +
\langle \hat{p}_j (0) \rangle  \kappa_j^{1/2}  \sin \omega_j t 
 \right ] \; \; .
\end{equation}

\noindent
It is convenient to rewrite the $c$-number equation (\ref{eqn5})
as follows;
\begin{equation}
\label{eqn7}
\langle \ddot{ \hat{X} } (t) \rangle + 
\int_0^t dt'  \beta (t-t')  \langle \dot{ \hat{X} } (t') \rangle
+ \langle V' ( \hat{X} ) \rangle
= F (t)
\end{equation}

\noindent
where we let the quantum mechanical mean value 
$ \langle \hat{F} (t) \rangle = F (t)$. We now turn to the {\it second}
average. To realize $F(t)$ as an effective 
$c$-number noise we now assume that the momenta 
$\langle \hat{p}_j (0) \rangle$ and the shifted co-ordinates
$\{ \langle \hat{q}_j (0) \rangle - \langle \hat{X} (0) \rangle \}$
of the bath oscillators are distributed according to a canonical distribution
of Gaussian forms as
\begin{equation}
\label{eqn8}
{\cal P}_j = {\cal N}
\exp \left \{ \frac{ -  [ \langle \hat{p}_j (0) \rangle^2 +
\kappa_j \left \{
\langle \hat{q}_j (0) \rangle - \langle \hat{X} (0) \rangle \right \}^2  
] }{
2 \hbar \omega_j \left ( \bar{n}_j + \frac{1}{2} \right ) }
\right \} 
\end{equation}

\noindent
so that for any quantum mechanical mean value
$O_j ( \langle\hat{p}_j (0) \rangle, 
\{ \langle \hat{q}_j (0) \rangle  - \langle \hat{X} (0) \rangle \} )$ the 
statistical average $\langle \ldots \rangle_S$ is
\begin{eqnarray}
\langle O_j \rangle_S & = & \int 
O_j ( \langle \hat{p}_j (0) \rangle, 
\{ \langle \hat{q}_j (0) \rangle  - \langle \hat{X} (0) \rangle \} )
\nonumber \\
& & \times  {\cal P}_j ( \langle \hat{p}_j (0) \rangle, 
\{ \langle \hat{q}_j (0) \rangle - \langle \hat{X} (0) \rangle \} ) 
\nonumber \\
& & \times d\langle \hat{p}_j (0) \rangle   
d \{ \langle \hat{q}_j (0) \rangle - \langle \hat{X} (0) \rangle \} \; \; .
\label{eqn9}
\end{eqnarray}

\noindent
Here $\bar{n}_j$ indicates the average thermal photon number of the $j$-th
oscillator at temperature $T$ and 
$\bar{n}_j = 1/[\exp \left ( \hbar \omega_j/k_BT \right ) - 1]$ and
${\cal N}$ is the normalization constant.

The distribution (\ref{eqn8}) and the definition of statistical average
(\ref{eqn9}) imply that $F(t)$ must satisfy
\begin{equation}
\label{eqn10}
\langle F (t) \rangle_S = 0
\end{equation}

\noindent
and
\begin{equation}
\label{eqn11}
\langle F (t) F (t') \rangle_S
= \frac{1}{2} \sum_j \kappa_j  \hbar \omega_j  
\left ( \coth \frac{\hbar\omega_j}{2k_BT} \right )  \cos \omega_j (t-t')
\; \; .
\end{equation}

\noindent
That is, the $c$-number noise $F(t)$ is such that it is zero centered and 
satisfies the standard quantum fluctuation-dissipation relation (FDR)
as known in the literature \cite{west} in
terms of quantum statistical average of the noise operators. 
The distribution (\ref{eqn8}) is thus an ansatz introduced to calculate
the ensemble average over the quantum-mechanical mean values of the bath
oscillators. Its justification lies in the fact that with (\ref{eqn9})
it reproduces the correct noise properties of the bath, i.e., the quantum 
fluctuation-dissipation relation for the $c$-number quantum noise 
(\ref{eqn11}) along with (\ref{eqn10}). Secondly, the distribution
(\ref{eqn8}) has a form which is Boltzmann-like (but not a Boltzmann
distribution) since the width parameter of the Boltzmann distribution
$ kT $ gets replaced by $ \hbar \omega_j ( \overline n_j + \frac {1}{2} ) $

To proceed further we now add the force term $V'(\langle \hat{x} \rangle )$
on both sides of Eq.(\ref{eqn7}) and rearrange it to obtain formally
\begin{equation}
\label{eqn12}
\ddot{x} (t) + \int_0^t dt'  \beta (t-t')  \dot{x} (t')
+ V' ( x )
= F (t) + Q (x, t)
\end{equation}

\noindent
where we let $\langle \hat{X} (t) \rangle = x (t)$ for simple notational
convenience and
\begin{equation}
\label{eqn13}
Q (x, t) = V' ( x ) - \langle V' (\hat{X} ) \rangle
\end{equation}

\noindent
represents the quantum mechanical dispersion of the force operator
$V'(\hat{X})$ due to the system degree of freedom. Since $Q(x,t)$ is a 
quantum fluctuation term Eq.(\ref{eqn12}) offers a simple interpretation.
This implies that the classical looking QGLE is governed by a $c$-number
quantum noise $F(t)$ which originates from the quantum mechanical
heat bath characterized by the properties (\ref{eqn10}) and
(\ref{eqn11}) and a quantum fluctuation term $Q(x,t)$ due to the quantum 
nature of the system characteristic of the nonlinearity of the potential. 
Although because of the last term in Eq.(\ref{eqn12}) the equation looks 
formal and implicit, the actual structure of $ Q(x,t) $ gets more transparent
as we go over to the beginning of the next section.
A recipe for calculation of $Q(x,t)$ is given in Ref.\cite{bbr,akp,sm}. 

We summarize the above discussions to point out that it is possible to 
formulate an exact QGLE (\ref{eqn12}) of the
quantum mechanical mean value of position of
a particle in a medium, provided the classical-like noise term $F(t)$
satisfies (\ref{eqn10}) and (\ref{eqn11}). The important new content
of the approach is that
to realize $F(t)$ as a noise term we have split up the standard
quantum statistical averaging procedure
into a quantum mechanical mean $\langle \ldots \rangle$ 
by explicitly using an initial coherent state representation of the bath 
oscillators and then a statistical average $\langle \ldots \rangle_S$
of the quantum mechanical mean values with distribution (\ref{eqn8}).
This is distinctly different from the usual procedure of quantum
statistical averaging where the quantum mechanical average is carried out
with number states over the noise operators followed by an ensemble
average with Boltzmann distribution.
Two pertinent points are to be noted:  First, it may be easily verified that
the distribution of quantum mechanical
mean values of the bath oscillators (\ref{eqn8})
reduces to classical Maxwell-Boltzmann
distribution in the thermal limit, $\hbar \omega_j \ll k_BT$.
Second, the vacuum term in the distribution (\ref{eqn8}) prevents the 
distribution of quantum mechanical mean values from being singular at
$T=0$; or in other words the width of distribution remains finite even 
at absolute zero, which is a simple consequence of uncertainty principle.
The procedure has been recently implemented by us to formulate a quantum
theory of Brownian motion \cite{bbr} and to propose an {\it exact}
non-Markovian quantum Kramers' equation \cite{dbbr}
with {\it true probability distribution functions}.


\section{Quantum Kramers' equation in energy space}

Let us begin by noting that the generalized quantum Langevin equation
(\ref{eqn12}) of a Brownian particle in presence of an external force field
takes into account of arbitrary coupling between the system and heat bath
and contains quantum corrections, $Q(x,t)$ due to system to all orders.
To make the later assertion explicit we now express the operators 
$ \hat{X} $ and $ \hat{P} $ as
\begin{mathletters}

\begin{eqnarray}
\hat X(t) & = &  \langle \hat X(t) \rangle + \delta \hat X (t) \nonumber \\
\hat P(t) & = &  \langle \hat P(t) \rangle + \delta \hat P (t) \; \; .
\label{eq6}
\end{eqnarray}
 
\noindent
By construction $ \langle \delta \hat X(t) \rangle = 0 $,  
  $ \langle \delta \hat P(t) \rangle = 0 $  and  
$ [\delta \hat X,\delta \hat P] = i\hbar $. 
Expanding $ \langle V^\prime (\hat X) \rangle $ around $ \langle \hat X
\rangle $ ($\equiv x$) in a Taylor series we obtain 

\begin{equation}
\langle V^\prime (\hat{X}) \rangle = V^\prime (x)\; +\; \frac {1}{2} 
V^{\prime \prime \prime} (x) \langle \delta {\hat X} ^2 \rangle\;+\; 
.\;.\;.   \label{eq7}
\end{equation}
 
\noindent
Therefore $Q(t)$ can be expressed as 

\begin{equation}
Q(t) = -  \sum_{n=3}^{\infty} \frac {1}{ (n - 1) ! } V^n (x) 
\langle \delta { \hat X  } ^{(n-1)} (t) \rangle \; \; .
\label{eq8}
\end{equation}

\noindent
Here $ V^n (x) $ denotes the n-th derivative of the classical potential.
The role of $ Q(x,t) $ is therefore to modify the classical potential
$ V(x) $ in Eq.(\ref{eqn12}). $ Q(x,t) $ can be calculated by solving
$ \langle \delta \hat X ^n (t) \rangle $ order by order. To the lowest 
order (second) 
$ \langle \hat X \rangle $ and $ \langle \delta {\hat X}^2 \rangle $
follow a coupled set of equations as given in Eqs.55(a-e) of 
Ref.\cite{bbr}.
(Higher order equations, e.g., the fourth order equations, are given in
Ref.\cite{sm}).
For convenience, we will now split up the right hand side of Eq.(\ref{eq8}) 
into a time-independent and a time-dependent part as,

\begin{equation}
\label{eq9}
Q (t) = - \sum _{n=3} ^{\infty} \frac {1}{ (n - 1)! } V^n (x)
\langle \delta {\hat X} ^{(n-1)} (0) \rangle \; + \; g(t)
\end{equation}

\noindent
where

\begin{equation}
\label{eq10}
g(t) = - \sum _{n=3} ^{\infty} \frac {1}{ (n - 1)! } V^n (x) 
\left [ \langle \delta {\hat X} ^{(n-1)} (t) \rangle\;-\; \langle \delta
{\hat X} ^{(n-1)} (0) \rangle \right ] \; \; .
\end{equation}

\noindent
For future use it is convenient to write $g(t)$ 
in the Taylor series of the form,

\begin{equation}
\label{eq11}
g(t) = - \sum _{m=0} ^{\infty} \sum _{n=3} ^{\infty} \frac{1}{ (n - 1)! } 
\frac{t ^m}{m!} \left [
\frac{ \partial^m }{ \partial t^m }  V^n 
\left [ x(t) \right ] \left \{ \langle \delta {\hat X} ^{(n-1)} (t) 
\rangle - \langle \delta {\hat X} ^{(n-1)} (0) \rangle \right \} \right ]
_{t=0} \; \; .
\end{equation}

\end{mathletters}

\noindent
The Langevin equation (\ref{eqn12}) then reduces to

\begin{equation}
\label{eq12}
\ddot {x} + \int _0 ^t d \tau \beta (t- \tau) \dot {x} (\tau) +
V^\prime (x) + \sum _{n=3} ^{\infty} \frac {1}{ (n - 1)! } V^n (x)
\langle \delta {\hat X} ^{(n-1)} (0) \rangle
= F (t) + g(t) \; \; .
\end{equation}

\noindent
Expressing

\begin{equation}
\label{eq13}
V_q (x) =  V (x) + \sum _{n=3} ^{\infty} \frac {1}{ (n - 1)! } 
V^{n-1} (x) \langle \delta {\hat X} ^{ (n - 1) } (0) \rangle
\end{equation}

\noindent
Eq.(\ref{eq12}) takes the form

\begin{equation}
\label{eq14}
\dot {x} =  v 
\end{equation}

\begin{mathletters}

\begin{equation}
\dot {v} + \int _0 ^t d \tau \beta (t - \tau) v (\tau) + V_q^\prime (x) 
=   F (t) + g (t)  \label{eq15}
\end{equation}

\noindent
Eq.(\ref{eq15}) is our starting Langevin equation. The 
potential $V_q (x)$ appearing in (\ref{eq13}) and (\ref{eq15})
is not the classical potential but a renormalized one with quantum
corrections. The damping memory kernel (\ref{eqn4}) is identified by the 
fluctuation-dissipation relation (\ref{eqn11}) by noting that in the
continuum limit
\begin{equation}
\label{fdr}
\langle F(t) F(0) \rangle_S = \frac{1}{2} 
\int_0^\infty d\omega \kappa (\omega) \rho (\omega) \hbar \omega
\left ( \coth \frac{\hbar \omega }{2k_BT} \right ) \cos \omega t
\equiv C (t)
\end{equation}

\noindent
and
\begin{equation}
\beta (t) = 
\int_0^\infty d\omega \kappa (\omega) \rho (\omega) \cos \omega t
\end{equation}

\noindent
in the Fourier domain can be related as
\begin{equation}
\label{eq2p8c}
\tilde{C}^c (\omega) = \frac{1}{2} \hbar \omega 
\left ( \coth \frac{\hbar \omega }{2k_BT} \right )
\tilde{\beta}^c (\omega)
\end{equation}

\end{mathletters}

\noindent
where $\tilde{C}^c (\omega)$ and $\tilde{\beta}^c (\omega)$ are the cosine
transforms of $C (t)$ and $\beta (t)$,
respectively.
For convenience we now express the Fourier transform of $\beta (t)$ as 
\begin{equation}
\label{eq16}
\tilde \beta _n (\omega) = \int _0 ^\infty dt \beta (t) \exp (-in \omega t)
\end{equation}

\noindent
We now consider the following time scales in the dynamics relevant for 
energy diffusion in the weak friction limit,

\begin{equation}
\label{eq17}
\gamma  \ll  1/ \tau _c  \ll  \omega
\end{equation}

\noindent
where $ \gamma $ is the friction arising due to interaction with the bath,
evaluated in the Markovian limit. $ \tau _c $ is the correlation time 
of the noise due to heat bath and
$ \omega $ is the linearized system frequency, which for a Brownian 
particle is assumed to be very high. This separation of time scales in
(\ref{eq17}) and casting of an operator Langevin equation in $c$-number
form (\ref{eq14}-\ref{eq15}) allow us to implement a classical method
for solving the problem of quantum energy diffusion.
Following the standard procedure one can transform 
Eqs.(\ref{eq14}-\ref{eq15}) 
to the action (J) and 
angle ($ \phi $) co-ordinates with the help of a Jacobian matrix as
\begin{eqnarray*}
\left (
\begin{array}{c}
\dot{J} \\
\dot{\phi}
\end{array}
\right ) 
& = & 
\left (
\begin{array}{cc}
-\frac{\partial v}{\partial \phi} & \frac{\partial x}{\partial \phi} \\
\frac{\partial v}{\partial J}     & -\frac{\partial x}{\partial J}
\end{array}
\right ) 
\left (
\begin{array}{c}
\dot{x} \\
\dot{v}
\end{array}
\right )  \nonumber \\
& = &
\left (
\begin{array}{cc}
-\frac{\partial v}{\partial \phi} & \frac{\partial x}{\partial \phi} \\
\frac{\partial v}{\partial J}     & -\frac{\partial x}{\partial J}
\end{array}
\right ) 
\left (
\begin{array}{c}
\frac{\partial H}{\partial v} \\
-\frac{\partial H}{\partial x} - \int_0^t d\tau \beta (t-\tau ) v (\tau )
+ F (t) + g(t)
\end{array}
\right ) \; \; .
\end{eqnarray*}

\noindent
Thus we have
\begin{eqnarray}
\label{eq18}
\dot J  & = &  \frac { \partial x }{ \partial \phi } \left [ - \int _0 ^t
d \tau \beta ( t - \tau ) v ( \tau ) + F (t) + g(t) \right ]    \\
\dot \phi & = &  \omega (J) -  \frac { \partial x }{ \partial J }
\left [ - \int _0 ^t d \tau \beta ( t - \tau ) v ( \tau ) + F (t) + g (t) 
\right ]   \label{eq19}
\end{eqnarray}

\noindent
Here $ v $ represents the velocity of the particle.
For the deterministic part of the system's Hamiltonian given by 
$  H = (1/2) v^2 + V_q (x)  $ we may write,

\begin{equation}
\label{eq20}
\omega (J)  =  \frac { d H (J) }{ d J } \; \; .
\end{equation}

\noindent
Since $V_q (x)$ (see Eq.(\ref{eq13})) contains quantum corrections, our
$J$ and $\phi$ are quantum ($c$-number) variables as implied in
(\ref{eq13}). In the absence of quantum corrections they become classical 
variables of CN \cite{cn}.
The canonical transformation from ($x, v$) space to
$ (J, \phi)  $ space has been done with the deterministic 
Hamiltonian. We can therefore expand $x$ and $v$ as,
\begin{mathletters}

\begin{eqnarray}
\label{eq21}
x ( J, \phi ) = \sum _{ n = -\infty } ^ \infty  x _n (J) \exp ( in \phi ) \\
v ( J, \phi ) = \sum _{ n = -\infty } ^ \infty  v _n (J) \exp ( in \phi )
\label{eq22}
\end{eqnarray}
\end{mathletters}

\noindent
with

\begin{equation}
\label{eq23}
x _n = x _{-n} ^* \; \; {\rm and}  \; \; v _n = v _{-n} ^* \; \; .
\end{equation}

\noindent
Differentiating  Eq.(\ref{eq21}) with respect to time and noting that in the
action-angle variable space  $ \dot \phi = \omega (J)  $  we can write,

\begin{equation}
\label{eq24}
v _n ( J ) = in \omega (J) x _n (J) \; \; .
\end{equation}

\noindent
Since, we are considering the motion in one dimension only, we can choose
$J$ and $ \phi $ in such a way that we can make the simplification
for $x = x^\star$ as,

\begin{eqnarray*}
x  =  \frac {1}{2} \sum _{n = -\infty} ^\infty
\left [ x _n \exp (in \phi) + x _n ^* \exp (-in \phi) \right ] \; \; .
\end{eqnarray*}

\noindent
Inserting Eq.(\ref{eq23}) we get,

\begin{eqnarray*}
x = \frac {1}{2} \sum _{ n = -\infty } ^\infty \left [ x _n \exp (in \phi)
 + x _{-n} \exp (-in \phi) \right ] \; \; .
\end{eqnarray*}

\noindent
With the choice of phase
\begin{mathletters}

\begin{equation}
\label{eq25}
x = x _{-n} \; \; \; \left [ \rm { since \; Im } (x _n) = 0 \right ]
\end{equation}

\noindent
$x$ may be further expressed as, 

\begin{eqnarray*}
x = \sum _{n = - \infty} ^{\infty} x _n \cos n \phi \; \; .
\end{eqnarray*}

\noindent
Similarly using Eq.(\ref{eq24}) and (\ref{eq25}) we get for
$v_n = - v_{-n}$

\begin{equation}
\label{eq26}
v = \sum _{n = -\infty} ^\infty v _n \sin n \phi \; \; .
\end{equation}

\end{mathletters}

\noindent
Inserting Eq.(\ref{eq21}) and (\ref{eq22}) in Eq.(\ref{eq18}) and 
(\ref{eq19}) we obtain,

\begin{eqnarray}
\label{eq27}
\dot J & = & - i \sum _{n = - \infty } ^{ \infty } \sum _{m = - \infty}
^{ \infty }n x _n \exp ( in \phi ) \int _0 ^t d \tau \beta (t - \tau) v _m 
\exp(im \phi) 
\nonumber  \\
& & + i S (t) \sum  _{n = - \infty} ^\infty n x _n \exp(in \phi) \\
\label{eq28}
\dot \phi & = & \omega (J) + \sum _{ n = - \infty } ^\infty 
\sum _{m = - \infty} ^{\infty} \frac { \partial x _n }{ \partial J } 
\exp(in \phi) \int _0 ^t d \tau \beta (t - \tau) v _m \exp(im \phi) 
\nonumber  \\
& & - S (t)\sum _{n = - \infty} ^\infty \frac { \partial x _n }{ \partial J } 
\exp(in \phi)
\end{eqnarray}

\noindent
where, we have expressed $S(t)$ as a sum of two terms; the noise due to
heat bath, $F(t)$ and quantum correction term, $g(t)$

\begin{equation}
\label{eq29}
S (t) = F (t) + g (t)
\end{equation}

\noindent
In the equations of motion (\ref{eq27}) and (\ref{eq28}), the argument
of the damping memory kernel $\beta$ is  $ (t - \tau) $. 
Now $\beta$ decays to zero in a time $ \tau_c $
(the correlation time). So, to deal with the integrals of 
Eq.(\ref{eq27}) and Eq.(\ref{eq28}), it is reasonable to divide the range of
integration into two parts: (a)  $  | t - \tau | \le \tau _c  $
and  (b) $  t \gg \tau _c  $. Thus following CN \cite{cn} we can write

\begin{eqnarray*}
\phi (t) & = & \phi [ \tau + ( t - \tau ) ]   \\
&  \simeq & \phi ( \tau ) + 
\left. \frac { \partial \phi }{ \partial t } 
\right  |_{t=\tau} ( t - \tau ) \; \; ,
\end{eqnarray*}

\noindent
neglecting higher terms of $  \tau _c  $. It follows that,

\begin{eqnarray}
\label{eq30}
\phi ( \tau ) & \simeq & \phi (t) - (t - \tau ) \omega  \\
\label{eq31}   
{\rm and} \; \; \; \; \; v _m (\tau)  & \simeq & v _m (t)
\end{eqnarray}

\noindent
Eq.(\ref{eq30}) and Eq.(\ref{eq31}) are reasonable approximations so far as
the integrals of Eq.(\ref{eq27}) and Eq.(\ref{eq28}) are concerned. Within the
integral, we therefore manipulate the behaviour of $ \phi $ and $ v _m $
for a time upto which $ \beta (t -\tau ) $ exists and also for the observational
time at which  $ \beta $ has decayed to zero. So, more specifically we can 
write for $ | t - \tau | \leq \tau _c $,

\begin{equation}
\label{eq32}
\int _0 ^t d\tau \beta ( t - \tau ) v _m ( \tau ) \exp [im \phi ( \tau )] \simeq
v _m (t) \exp [im \phi (t)] \int _0 ^t d\tau \beta (t - \tau) 
\exp [-im (t - \tau) \omega]
\end{equation}

\noindent
and for $ t \gg \tau _c $, using (\ref{eq16}) we have

\begin{equation}
\label{eq33}
\int _0 ^t d\tau \beta (t - \tau) v _m (\tau) \exp [im \phi (\tau )]  \simeq
v_m (t) \exp [ im \phi (t) ] \tilde{\beta}_m (\omega) \; \; .
\end{equation}

\noindent
Putting Eq.(\ref{eq33}) which takes into account the observational time 
scale, in Eq.(\ref{eq27}) and Eq.(\ref{eq28}) we get,

\begin{eqnarray}
\label{eq34}
\dot J & = & -i \sum _{n = -\infty} ^{\infty} \sum _{m = -\infty} ^{\infty}
n x _n v _m \tilde {\beta} _m (\omega) \exp [i (n + m) \phi] + 
i S (t) \sum _{n = -\infty} ^{\infty} n x _n \exp (in \phi)   
\\
\label{eq35}
{\rm and} \; \; 
\dot \phi & = & \omega (J) + \sum _{n = -\infty} ^\infty \sum _{m = -\infty} 
^\infty x _n ^\prime v _m \tilde {\beta} _m (\omega) \exp [i (n + m) \phi] -  
S (t) \sum _{n = -\infty} ^\infty x _n ^\prime \exp (in \phi)   
\\
\label{eq36}
{\rm where} \; \; 
x _n ^\prime & = & \frac { \partial x _n }{ \partial J } \; \; .
\end{eqnarray}

\noindent
Our next task is to formulate the Fokker-Planck equation. 
To this end we note that Lax \cite{lax} had prescribed a method for deriving
Markovian Fokker-Planck equation from a classical Langevin equation
with short but finite correlation time. Although the procedure can be 
extended to higher order iteration scheme to include non-Markovian effects
we adopt the method advocated by Carmeli and Nitzan \cite{cn} for their
classical theory. This is based on Kramers-Moyal expansion of the transition
probability which connects the 
probability distribution function $ P ( J, \phi, t ) $ at time  $t$ with that
of $ P (J, \phi, t+\tau) $ at a later time $ t + \tau $ for small $ \tau $,
given that we know the moments of the distribution. For details we refer to
Risken \cite{risken}
The time evolution of the probability distribution $ P ( J, \phi, t ) $ is 
determined by the equation,

\begin{equation}
\label{eq37}
\frac { \partial P}{ \partial t } = 
\lim _{\tau \rightarrow 0+} \left [ \frac {1 }{ \tau } \sum  _{n = 1} ^ \infty
\frac { {(-1) ^n} }{ n! }  \sum_{(m,k = 0); (m + k = n)}
{ \left ( \frac { \partial }{ \partial J } \right ) } ^m 
{ \left ( \frac { \partial }{ \partial \phi } \right ) } ^k 
\left \{ \langle { (\Delta J _t) } ^m { (\Delta \phi _t) } ^k \rangle_S  P  
\right \} \right ]
\end{equation}

\noindent
where

\begin{eqnarray*}
\Delta J _t & = & \Delta J _t (\tau) = J (t + \tau) - J (t)  \\
\Delta \phi _t & = & \Delta \phi _t (\tau) = \phi (t + \tau) - \phi (t)
\; \; .
\end{eqnarray*}

\noindent
At this juncture it is worth recalling that $ \tau $ is the coarse-grained
time scale over which the probability distribution function evolves, 
whereas $ \tau _c $ is the correlation time,
which due to low damping is much smaller than $ \tau $.
The low value of $ \gamma $ prompts us to take 
$ { \gamma } ^ {-1} $ as the largest time scale for the entire problem. On
the other hand, the reciprocal of the frequency of oscillation, i.e.,
$ {\omega} ^{-1} $, is the smallest time scale. Our task is 
therefore to evaluate the moments of the form 
$ \langle { ( \Delta J _t ) } ^m { ( \Delta \phi _t ) } ^k \rangle_S $ where
our definition of average $\langle \ldots \rangle_S$ is given in
(\ref{eqn9}).

To evaluate the moments we make use of the following standard procedure 
\cite{cn,lax}:

\begin{eqnarray}
\label{eq38}
\Delta J_t (\tau) & = & \int _0 ^ \tau ds \dot J [ J (t + s), \phi (t + s),
t + s ]  \\
\label{eq39}
\Delta \phi_t (\tau) & = & \int _0 ^\tau ds \dot \phi [ J (t + s), 
\phi (t + s), (t + s) ]
\end{eqnarray}

\noindent
where the forms $ \dot J $ and $ \dot \phi $ are given by Eq.(\ref{eq34})
and (\ref{eq35}), respectively. The iterative equations are given by,

\begin{eqnarray}
\label{eq40}
\Delta J _t ^{ (l) } ( \tau ) & = & \int _0 ^\tau ds \dot J 
[ J (t) + 
\Delta J _t ^ { (l-1) } (s) , \phi (t) + \Delta \phi _t ^ { (l-1) } (s) ,
t + s ]   \\
\label{eq41}
\Delta \phi _t ^ { (l) } ( \tau ) & = & \int _0 ^ \tau ds \dot \phi
[ J (t) + 
\Delta J _t ^ { (l-1) } (s) , \phi (t) + \Delta \phi _t ^ { (l-1) } (s) ,
t + s ]   
\end{eqnarray}

\noindent
where $ (l) $ denotes the $ l ^ {\rm th} $ iteration stage.

The non-Markovian nature (i.e., $ \tau_c $ is finite and 
$ \tau _c < \tau $ ) of the present problem allows us to consider, in
principle, all orders of $ \tau $ in Eq.(\ref{eq37}). But, since 
$ { \partial P } / { \partial t } $ is evaluated in the limit 
$ \tau \rightarrow 0_+ $, terms linear in $ \tau $, i.e., the coarse-grained
time scale, are taken while all the higher powers are neglected. 
We now introduce the following abbreviations 

\begin{eqnarray}
\label{eq42}
\sigma _n (J) & = & in x_n (J) \\
\label{eq43}
\mu _n (J) & = & \frac { dx _n (J) }{ dJ }  \\
\label{eq44}
B _{nm} (J) & = & in x _n (J) v_m (J) \tilde {\beta}_m [ \omega (J) ]  \\
\label{eq45}
C _{nm} (J) & = & \left [ \frac { d x _n (J) }{ dJ } \right ] v _m (J) 
\tilde {\beta} _m [ \omega (J) ]
\end{eqnarray}

\noindent
Substituting Eq.(\ref{eq42})-(\ref{eq45}) in Eq.(\ref{eq34}) and (\ref{eq35}) 
we obtain the quantum equations in the form of classical equations of
CN \cite{cn}:

\begin{eqnarray}
\label{eq46}
\dot J & = & - \sum _{n = -\infty} ^ \infty \sum _{m = -\infty } ^\infty
B _{nm} (J) \exp [i (n + m) \phi] + S (t) \sum _{n = -\infty} ^\infty 
\sigma _n (J) \exp (in \phi)  \\  {\rm and} \; \; \;
\label{eq47}
\dot \phi & = & \omega (J) + \sum _{n = -\infty} ^{\infty} 
\sum _{m = -\infty} ^\infty C _{nm} (J) \exp [i (n + m) \phi] - 
S (t) \sum _{n = -\infty} ^\infty \mu _n (J) \exp (in \phi)
\end{eqnarray}

\noindent
From Eq.(\ref{eq40}) and (\ref{eq46}) we get the explicit structure of  
$ \Delta J _t $ as, 

\begin{eqnarray}
\label{eq48}
\Delta J _t (\tau) = 
& - & \sum _{n = -\infty} ^\infty \sum _ {m = -\infty} ^\infty \int _0 ^\tau
ds B _{nm} [ J (t) + \Delta J _t (s) ] \exp \{ i (n + m) [ \phi (t) + 
\Delta \phi _t (s) ] \}  \nonumber  \\  
& + & \sum _{n = -\infty} ^\infty \int _0 ^\tau ds 
S (s) \sigma _n [ J (t) + \Delta J _t (s) ] \exp \{ in [ \phi (t) + 
\Delta \phi _t (s) ] \} \; \; .
\end{eqnarray}

\noindent
Similarly from Eq.(\ref{eq41}) and (\ref{eq47}), $ \Delta \phi _t $ is given 
by,

\begin{eqnarray}
\label{eq49}
\Delta \phi _t (\tau) & = &
\int _0 ^\tau ds \omega [ J (t) + \Delta J _t (s) ]  \nonumber  \\
& + & \sum _{n = -\infty} ^\infty \sum _{m = -\infty} ^\infty \int_ 0 ^\tau
ds C _{nm} [ J (t) + \Delta J _t (s) ] \exp \{ i (n + m) [ \phi (t)
+ \Delta \phi _t (s) ] \}   \nonumber  \\
& - & \sum _{n = -\infty} ^\infty \int _0 ^\tau ds S (s) \mu _n [ J (t)
+ \Delta J _t (s) ] \exp \{ in [ \phi (t) + \Delta \phi _t (s) ] \}
\; \; .
\end{eqnarray}

\noindent
For beginning the systematic iteration procedure given by Eq.(\ref{eq48})
and (\ref{eq49}) we initialize the zero order iteration stage as,

\begin{mathletters}

\begin{eqnarray}
\label{eq50}
\Delta J _t ^{ (0) } (\tau) & = & 0  \\    {\rm and} \; \; \; \;
\label{eq51}
\Delta \phi _t ^{ (0) } (\tau) & = & \omega [ J (t) ] \tau = \omega \tau
\; \; .
\end{eqnarray}

\end{mathletters}

\noindent
The entire process of iteration involves cumbersome but straightforward
calculations, some relevant details of which appear in the Appendix-A. 
Here we state only the main results.

Inserting Eqs.(\ref{eq50}) and (\ref{eq51}) in the right hand side of 
Eqs.(\ref{eq48}) and (\ref{eq49}) we get the results of the first order 
iteration. Thus,

\begin{eqnarray}
\label{eq52}
\Delta J _t ^{ (1) } (\tau) & = & 
- \tau \sum _{n = -\infty} ^\infty B _{ n,-n } + 
\sum _{n = -\infty} ^\infty \sigma _n \exp (in \phi) \int _0 ^\tau ds F (s) 
\exp (in \omega s)  \nonumber  \\
& + & \sum _{n = -\infty} ^\infty \sigma _n \exp (in \phi)
\int _0 ^\tau ds g (s) \exp (in \omega s)
\end{eqnarray}

\noindent
and

\begin{eqnarray}
\label{eq53}
\Delta \phi _t ^{ (1) } (\tau) & = &   
\omega \tau + \tau \sum _{n = -\infty} ^\infty C _{n,-n} - 
\sum _{n = -\infty} ^\infty \mu _n \exp (in \phi) \int _0 ^\tau ds F (s)
\exp (in \omega s)  \nonumber \\
& - & \sum _{n = -\infty} ^\infty \mu _n \exp (in \phi) \int _0 ^\tau ds
g (s) \exp (in \omega s)
\end{eqnarray}

\noindent
where in writing Eq.(\ref{eq52}) and (\ref{eq53}) we have used 
Eq.(\ref{eq29}). For the second iteration we put Eq.(\ref{eq52}) and 
(\ref{eq53}) back into Eq.(\ref{eq48}) and (\ref{eq49}) and thus obtain 
$ \Delta J _t ^{(2)} (\tau) $ and $ \Delta \phi _t ^ {(2)} (\tau) $. Putting
them back into Eq.(\ref{eq48}) and (\ref{eq49}) again we obtain 
$ \Delta J _t ^{(3)} (\tau) $ and $ \Delta \phi _t ^ {(3)} (\tau) $. These
are presented in some details in the Appendix.

In calculating the moments as demanded by Eq.(\ref{eq37}), we have neglected
all higher powers ($ n \ge 2 $) of $ \tau $ and $ 1/ \omega $. 
The reason for doing this in case of $ \tau $ is clear from the limit  
imposed on $ \tau $ in Eq.(\ref{eq37}). For $ 1/ \omega $ also, this 
approximation is legitimate since $ 1/ \omega $ is the shortest time 
scale of the problem (see inequality (\ref{eq17})). 
The final results for the moments are:

\begin{eqnarray}
\label{eq54}
\langle [ \Delta J_t (\tau) ]^2 \rangle_S & = & 4 \tau 
\sum_{n=1}^\infty n^2 |x_n|^2 \tilde{C}_n^c (\omega) \\
\label{eq55}
\langle [ \Delta \phi_t (\tau) ]^2 \rangle_S & = & 4 \tau 
\sum_{n=1}^\infty   \left | \frac{dx_n}{dJ} \right |^2
\tilde{C}_n^c (\omega) \\
\label{eq56}
\langle [ \Delta J_t (\tau) ][ \Delta \phi_t (\tau) ] \rangle_S & = & 0 \\
\label{eq57}
\langle \Delta J_t (\tau) \rangle_S & = & - 2 \tau
\sum_{n=1}^\infty n^2 \left [ \omega |x_n|^2 \tilde{\beta}_n^c (\omega) -
\frac{d}{dJ} \left \{ |x_n|^2 \tilde{C}_n^c (\omega) \right \} \right ] \\
\label{eq58}
\langle \Delta \phi_t (\tau) \rangle_S & = & \omega \tau + \tau
\sum_{n=1}^\infty n  \left [ \omega \tilde{\beta}_n^s
\frac{d |x_n|^2}{dJ} - \frac{d}{dJ} \left ( \tilde{C}_n^s
\frac{d |x_n|^2}{dJ} \right ) \right ] - \tau f_0^\prime \mu_0 t_c
\end{eqnarray}

\noindent
where
\begin{mathletters}

\begin{eqnarray}
\label{eq59a}
\tilde{\beta}_n^c & = &  \int_0^\infty dt \beta (t) \cos (n\omega t) \\
\label{eq59b}
\tilde{\beta}_n^s & = &  \int_0^\infty dt \beta (t) \sin (n\omega t) \\
\label{eq59c}
\tilde{C}_n^c & = &  \int_0^\infty dt C (t) \cos (n\omega t) \\
\label{eq59d}
\tilde{C}_n^s & = &  \int_0^\infty dt C (t) \sin (n\omega t) 
\end{eqnarray}

\noindent
and
\begin{equation}
\label{eq59e}
f_0^\prime = - \sum_{n=3}^\infty \frac{1}{(n-1)!} 
\frac{\partial}{\partial t} \left [ V^n [x(t)] 
\left \{ \langle \delta \hat{X}^{n-1} (t) \rangle -
\langle \delta \hat{X}^{n-1} (0) \rangle \right \} \right ] \; \; .
\end{equation}

\noindent
Also

\begin{eqnarray}
\label{eq59f}
\tilde{\beta}_n (\omega) & = & \tilde{\beta}_n^c (\omega) - i \tilde{\beta}_n^s (\omega) \\
\label{eq59g}
\tilde{C}_n (\omega) & = & \tilde{C}_n^c (\omega) - i \tilde{C}_n^s (\omega)
\end{eqnarray}

\end{mathletters}

\noindent
Some remarks are needed in connection with Eq.(\ref{eq54}) to
Eq.(\ref{eq58}). 
Let us now examine how the quantum notion is implied in 
Eqs.(\ref{eq54}-\ref{eq58}). First, all the moments are the functions of the 
Fourier components $x_n (J)$ where $J$ is a quantum $c$-number. Second,
the moments are crucially dependent on the Fourier components of quantum
correlation function $C(t)$ of the heat bath. In the classical limit
$x_n (J)$ becomes the functions of the classical action variable $J$ and also
$\tilde{C}_n (\omega )$ reduces to
$\tilde{C}_n (\omega ) = k_BT \tilde{\beta}_n (\omega )$. We thus obtain,

\begin{eqnarray}
\langle [ \Delta J_t (\tau) ]^2 \rangle_S & = &  4 \tau k_B T \sum_{n = 1}
^\infty n^2 | x_n |^2 {\tilde \beta} _n ^c (\omega)  \nonumber  \\
\langle [ \Delta \phi _t (\tau) ]^2 \rangle_S & = & 4 \tau k_B T \sum_{n = 1} 
^\infty \left | \frac { dx_n }{ dJ } \right |^2 {\tilde \beta}_n^c (\omega)
\end{eqnarray}

\noindent
in the high-temperature limit $ \hbar \omega \ll k_B T $. 
The last term in Eq.(\ref{eq58}) 
is due to a correction to frequency $\omega$ and is of pure quantum
origin (\ref{eq11}).
$ f_0^\prime $ is precisely the co-efficient of $ t $ in the
Taylor expansion of Eq.(\ref{eq11}). $ t_c $ is the `cut-off' time upto
which the quantum fluctuation remains linear in time and is approximated as
$\sim 1/ \omega $, as allowed by the time scale of the problem.
Thus the quantum character of the nonlinear system enters into the 
description in two different ways. First, classical Hamiltonian gets
modified by quantum corrections at $t=0$ (see Eq.(\ref{eq13})). This
makes action angle variables bear quantum signature. Second, the quantum
correction for $t > 0$ as contained in $g(t)$ makes its presence in 
phase drift term in (\ref{eq58}).

Inserting Eq.(\ref{eq54}) to (\ref{eq58}) in Eq.(\ref{eq37}) and thereby
disregarding terms with $n > 2$  with the following definitions

\begin{eqnarray}
\label{eq60}
\epsilon (J) & = & 2 \sum_{n = 1} ^\infty n^2 | x_n |^2 
\tilde \beta_n^c (\omega)  \\
\label{eq61}
\Gamma (J) & = & 2 \sum_{n = 1} ^\infty \left | \frac { dx_n }{ dJ } \right |
^2 \tilde C _n^c (\omega)   \\
\label{eq62}
\Omega(J) & = & \omega + \sum_{n = 1} ^\infty n \left [ \omega \tilde \beta_n^s
\frac { d | x_n |^2 }{ dJ } - \frac { d }{ dJ } \left ( \tilde C_n^s
\frac { d | x_n |^2 }{ dJ } \right ) \right ] - f_0^\prime \mu_0 t_c
\end{eqnarray}

\noindent
we obtain the Fokker-Planck equation for $ P ( J, \phi,t ) $ as, 

\begin{eqnarray}
\label{eq63}
\frac { \partial P ( J,\phi,t )  }{ \partial t } & = &
\frac { \partial }{ \partial J } 
\left [ 2 \sum _{n = 1}^\infty n^2 | x_n |^2 
\tilde \beta_n^c (\omega) \left \{ \frac {\tilde C_n^c (\omega)}{\tilde \beta_n^c(\omega)} 
\frac { \partial }{ \partial J } + \omega (J) \right \} P \right ]
\nonumber   \\
& + & \Gamma (J) \frac { \partial^2 P }{ \partial \phi^2 } - \Omega (J)
\frac { \partial P }{ \partial \phi } \; \; .
\end{eqnarray}

\noindent
If the distribution function is initially independent of $\phi$ it satisfies
the quantum diffusion equation in action space

\begin{equation}
\label{eq64}
\frac { \partial P ( J,t )  }{ \partial t } = 
\frac { \partial }{ \partial J } \left [ \epsilon (J) \left \{ \Lambda
\frac { \partial }{ \partial J } + \omega (J) \right \} P \right ]
\end{equation}

\noindent
where by virtue of (\ref{eq2p8c}) we write

\begin{eqnarray}
\Lambda & = & \Lambda ( \overline {\omega} ) \simeq 
\frac { \tilde C_n^c ( \overline {\omega} ) }
{ \tilde \beta_n^c(\overline {\omega} ) }  
\nonumber    \\   {\rm or} \; \; \; \; 
\label{eq65}
\Lambda & = & \hbar \overline {\omega} \; [ \overline {n} 
( \overline {\omega} ) +  1/2 ]
\end{eqnarray}

\noindent
Here $ \overline {\omega} $ is the linearized frequency and $\Lambda$
plays the typical role of $k_BT$. We have
\begin{eqnarray*}
\omega (J) = \frac{ \partial H }{ \partial J } = \frac{ dE }{ dJ }
\; \; .
\end{eqnarray*}

\noindent
Expressing

\begin{equation}
\label{eq66}
\omega (J) = \nu (E)
\end{equation}

\noindent
we have

\begin{equation}
\label{eq67}  
\frac{\partial}{ \partial J } = \nu (E) \frac{ \partial }{ \partial E }
\; \; .
\end{equation}

\noindent
With this transformation the quantum Kramers' equation for energy
diffusion [Eq.(\ref{eq64})] looks like,

\begin{equation}
\label{eq68}
\frac { \partial P ( E,t ) }{ \partial t } = \frac { \partial }{ \partial E }
\left [ D (E) \left ( \frac { \partial }{ \partial E } + 
\frac { 1 }{ \Lambda } \right ) \nu (E) P ( E,t ) \right ]
\end{equation}

\noindent
where the diffusion coefficient is given by

\begin{equation}
\label{eq69}
D (E) = \nu (E) 2 \hbar \overline{\omega} 
\left [ \overline{n} ( \overline{\omega} ) + \frac{1}{2} \right ] 
\sum_{n = 1}^\infty n^2 | x_n |^2
\int_0^\infty dt \beta (t) \cos [n \nu (E) t ] \; \; .
\end{equation}

\noindent
{\it Eq.(\ref{eq68}) is the first key result of the present paper}.
The equation is valid for arbitrary temperature and noise correlation.
The prime quantities that determine the equation for energy diffusion
(\ref{eq68}) are the diffusion coefficients $D$, the quantum analogue 
of $kT$, $\Lambda$ and the frequency of the dynamical system, $\nu(E)$.
It is important to note that all the quantities as defined by (\ref{eq69}),
(\ref{eq65}) and (\ref{eq66}), respectively contain quantum contributions. 
In the classical limit Eq.(\ref{eq65}) reduces to $kT$ when 
$ \overline{n} (\overline{\omega}) \gg 1/2 $ and 
$ \overline{n} (\overline{\omega}) \;
( = [ \exp(\hbar\overline{\omega}/kT) - 1 ]^{-1} )
\approx kT/\hbar\overline{\omega} $. Since by virtue (\ref{eq66})
$ \nu(E) = \omega(J) = \partial{H}/\partial{J} $ with $H$ defined as
$ H = (1/2) v^2 + V_q (x) $  where $ V_q (x) $ includes quantum corrections
over the classical potential $V(x)$ according to Eq.(\ref{eq13}), 
$\nu(E)$ reduces to classical frequency in the classical limit as
$ \hbar \rightarrow 0 $. Although the expression for the diffusion 
coefficient (\ref{eq69}) looks a bit complicated and formal due to the 
appearence of the Fourier coefficients $x_n$ in the summation, it is possible
to read the various terms in $D(E)$ in the following way. 
$D(E)$ is essentially an approximate product of three terms, 
$ \hbar \overline{\omega} [\overline{n}(\overline{\omega}) + 1/2] $,
$ \int_0^{\infty} dt \beta(t) \cos [n\nu(E)t] $ and 
$ \nu(E) \sum_{n=1}^{\infty} n^2 \mid x_n \mid^2 $, where the $n$ dependence of 
the latter two contributions have been separated out for interpretation.
The integral is the Fourier transform of the memory kernel, while the
sum can be shown to be equal to $J$ ( Appendix D of Ref.\cite{cn} ), which is 
the quantum action variable. In the classical limit the quantum diffusion
coefficient $D(E)$ therefore clearly reduces to the classical diffusion
coefficient of Carmeli and Nitzan \cite{cn}.
A few further remarks on the related issues may be made at this point. To
address the problem of nonequilibrium quantum tunneling above cross-over
temperature (nonequilibrium situation arises due to the significant
growth of population above zero levels at temperature above cross-over
since the dissipation is very weak), several authors
\cite{vm,lo,rj,chow,dekker,griff} have advocated the use of a probability
function per unit time (of finding the system in the barrier region near
a classical turning point with energy $E$) which obeys an integral
equation \cite{griff} whose the differential approximation leads to an
equation similar (not the same) to Eq.(\ref{eq68}). The notable difference
is in the fact that the former equation 
is applicable above cross-over temperature
while Eq.(\ref{eq68}) works at all temperature down to vacuum limit. The
energy loss coefficient in (\ref{eq68}), i.e., $D(E)/\Lambda$ when put into
the form $[$ using (\ref{eq65}) $]$,
\begin{eqnarray*}
\frac{D(E)}{\Lambda} =
\int_0^\infty dt \beta (t) \nu (E) \sum_{n=1}^\infty 2 n^2
|x_n|^2 \cos [ n \nu (E) t ]
\end{eqnarray*}

\noindent
is comparable to that of Griff {\it et al}\cite{griff}, 
the integrand without $\beta (t)$ being a function of action 
(or equivalently energy). These results can be utilized as a consistency
check of the present scheme.


\section{Quantum Energy Diffusion Controlled Rate of Escape}

The classical treatment of memory effects in the energy diffusion controlled
escape is now well-documented in the literature \cite{cn,hw,gh1}. 
To address the corresponding problem in the quantum domain we start by
recasting the Kramers' equation in the  energy diffusion regime 
[Eq.(\ref{eq64})] in the form of a continuity equation to obtain,

\begin{equation}
\label{eq70}
\frac { \partial P (E,t) }{ \partial t } + 
\frac { \partial j_E }{ \partial E } = 0
\end{equation}

\noindent
where $ j_E $ is the flux along the energy co-ordinate at thermal 
equilibrium and is given by,

\begin{equation}
\label{eq71}
j_E = - D (E) \left [ \frac { \partial }{ \partial E } + \frac {1}{\Lambda}  
\right ] \nu (E) P_ {st} (E)
\end{equation}

\noindent 
where $ P_{st} $ is the stationary probability distribution.
\noindent 
For zero current condition, we have the equilibrium distribution,
$ P_{eq} $ at the source well as,

\begin{eqnarray}
P_{eq} (E) & = & \frac { N^{-1} }{ \nu (E) } \exp ( -E / \Lambda ) 
\nonumber \\
\label{eq72} 
& = & \frac { N^{-1} }{ \nu (E) } \exp [ -(E^c + \Delta Q) / \Lambda ]
\end{eqnarray}

\noindent
where we have split the energy into classical ( $ E^c $ )
and quantum ( $ \Delta Q $ ) parts, the contribution arising from the 
latter being very small. We now define the rate of escape $ k $ as 
flux over population

\begin{equation}
\label{eq73}
k = \frac { j_E }{ n_a }
\end{equation}

\noindent
where

\begin{eqnarray}
n_a & = & {\rm total\; population\; at\; the\; source\; well }  \nonumber  \\
\label{eq74}
& = & \int _0 ^{E_b^c} P (E) dE
\end{eqnarray}

\noindent
Here $ E_b^c $ is the classical value of the activation barrier. Following
BHL \cite{bhl} we use a Kramers' like ansatz 

\begin{equation}
\label{eq75}
P (E) = \eta (E) P_{eq} (E)
\end{equation}

\noindent
to arrive at

\begin{equation}
\label{eq76}
j_E = - D (E) \nu (E) P_{eq} (E) \frac { \partial \eta (E) }{ \partial E } 
\end{equation}

\noindent
Integrating the above expression from $ E = E_1 \simeq \Lambda $ 
(see Eq.(\ref{eq65}) ) to 
$ E = E_b^c $, one derives an expression for energy independent current  
$ j_E $ (with $ E \le E_b^c $) as, 

\begin{eqnarray}
j_E & = & \frac { [ \eta (\Lambda) - \eta (E_b^c) ] }{ \int_\Lambda^{E_b^c}   
\frac { dE }{ D (E) \nu (E) P_{eq} (E) } }  \nonumber   \\
\label{eq77}
& = & [ 1 - \eta (E_b^c) ] D (E_b^c) \frac { N^{-1} }{ \Lambda } 
e^{ - E_b^c / \Lambda }
\end{eqnarray}

\noindent
where we have used the boundary condition $ \eta (\Lambda) \simeq 1 $.

Following the original reasoning by BHL we now allow an outflow $ j_{out} $
from each energy range $ E $ to $ E + dE $, with each $ E $ satisfying
the condition $ E \ge E_b^c $. Then we can write,

\begin{equation}
\label{eq78}
dj_{out} = \alpha \nu (E) \eta (E) P_{eq} (E) dE
\end{equation}

\noindent
which is compensated by a divergence in the vertical flow,

\begin{equation}
\label{eq79}
\frac { dj_E }{ dE } = - \alpha \nu (E) \eta (E) P_{eq} (E) 
\end{equation}

\noindent
Here $ \alpha $ is a parameter which has been set approximately equal to one
by BHL. Inserting the expression for non-equilibrium current Eq.(\ref{eq76}),
we obtain the ordinary differential equation for $ \eta (E) $ as,

\begin{equation}
\label{eq80}
D (E) \frac { d^2 \eta }{ d E^2 } + \left [ \frac { dD (E) }{ dE } - 
D (E) \frac {1}{ \Lambda } \right ] \frac { d\eta }{ dE } - \alpha \eta (E)
= 0 \; \; .
\end{equation}

\noindent
Within small energy range above $ E_b^c $ one can assume essentially a 
constant diffusion co-efficient i.e.,

\begin{equation}
\left [ \frac { dD (E) }{ dE } \right ]_{ E \simeq E_b^c } = 0
\; \; {\rm for} \; E \ge E_b^c \; \; .
\end{equation}

\noindent
Substituting a trial solution of the form 
$ \eta (E) = C \exp ( sE / \Lambda ) $ for $ s < 0 $, in Eq.(\ref{eq80})
we have,

\begin{equation}
\label{eq81}
s_- = - \frac {1}{2} \left [ \left ( 1 + \frac { 4 \alpha \Lambda^2 }
{ D(E_b^c) } \right )^{ 1/2 } - 1 \right ] \; \; .
\end{equation}

\noindent
Setting $ \eta (E) = \eta ( E_b^c ) \exp [ s (E - E_b^c) / \Lambda ] $ and 
putting this into Eq.(\ref{eq76}) and comparing this with the right hand side of 
Eq.(\ref{eq77}) we have ,

\begin{equation}
\label{eq82}
\eta ( E_b^c ) = 1 / (1 - s)   \; \; \; {\rm for}\; s < 0
\end{equation}

\noindent
Thus, escape rate $ k $ can be obtained as,

\begin{equation}
\label{eq83}
k = j_E \left [ \int_0^{E_b^c} \eta (E) P_{eq} (E) dE \right ]^{-1} \; \; .
\end{equation}

\noindent
Making use of (\ref{eq82}) in (\ref{eq77}) and the resulting expression
for $j$ in (\ref{eq83}) we obtain

\begin{equation}
\label{eq84}
k = \frac { -s }{ 1 - s } \left [ 
\frac{ \int_0^{E_b^c} \eta (E) P_{eq} (E) dE 
}{ 
(N^{-1} / \Lambda) D (E_b^c) \exp ( -E_b^c / \Lambda ) } 
\right ]^{-1} \; \; .
\end{equation}

\noindent
For the dynamics at the bottom we have $ \eta \rightarrow 1 $.
Recalling that $ E = E^c + \Delta Q $, where $ \Delta Q $ is the
quantum contribution to classical energy, we expand $E$ in a Taylor series. 
Here $E^c = (v^2/2) + V(x) $ and $\Delta Q (x)$ is the quantum correction
terms in (\ref{eq13}). Retaining terms upto the   
second order in $ x $, and making harmonic approximation around the bottom 
of the well at $ x = 0 $, we get 

\begin{equation}
\label{eq86}
E = \frac {p^2}{2} + \frac {1}{2} \omega_0^2 x^2 + \Delta Q_0
+ \Delta Q_0^\prime x + \frac{1}{2} \Delta Q_0^{\prime\prime} x^2
\end{equation}

\noindent
where $\omega_0$ corresponds to the frequency at the bottom of the
classical potential $V(x)$ at $x=0$ so that 
$\omega_0^2 = \partial^2 V(x)/ \partial x^2 |_{x=0}$. The subscript
zero in $\Delta Q_0$, $\Delta Q_0'$ and $\Delta Q_0''$ are the quantities
evaluated at this point.

Now $ n_a $, the total population at the source well, can be evaluated as,

\begin{eqnarray*}
n_a &=& \int_{-\infty}^\infty \int_{-\infty}^\infty P_{eq} (E) dxdp\\
    &=& \exp[ -\Delta Q_0 / \Lambda ] 
\int_{ -\infty }^\infty \exp(-v^2 / 2\Lambda) dv
\int_{ -\infty }^\infty \exp \left [ -\frac {1}{\Lambda} \left (
\frac {1}{2} \omega_0^2 x^2 + \Delta Q_0^\prime x + \frac {1}{2} \Delta 
Q_0^{\prime\prime} x^2 \right ) \right ] dx \; \; .
\end{eqnarray*}

\noindent
Thus
\begin{equation}
\label{eq87}
n_a = \frac {1}{N} \frac { 2\pi \Lambda }
{ \sqrt {\omega_0^2 + \Delta Q_0^{\prime\prime} } }
\exp \left [ - \frac { \Delta Q_0 }{ \Lambda } + 
\frac { ( \Delta Q_0^\prime )^2 }
{ 2\Lambda ( \omega_0^2 + \Delta Q_0^{\prime\prime} ) } \right ]
\; \; .
\end{equation}

\noindent
So, the quantum non-Markovian rate of escape from a metastable well in the 
low-friction regime is given by,

\begin{eqnarray}
k & = & 
\left [ \frac { \{ 1 + (4 \alpha \Lambda^2) / D (E_b^c) \}^{1/2} - 1 }
{ \{ 1 + (4 \alpha \Lambda^2) / D (E_b^c) \}^{1/2} + 1 } \right ]
\frac { D (E_b^c) }{ \Lambda^2 } 
\frac { \sqrt {\omega_0^2 + \Delta Q_0^{\prime\prime} } }{ 2 \pi }
\nonumber \\
& & \times
\exp \left [ -\frac {1}{\Lambda} \left \{ E_b^c - \Delta Q_0 +
\frac { ( \Delta Q_0^\prime )^2 }{ 2( \omega_0^2 + 
\Delta Q_0^{\prime\prime} ) } \right \} \right ] \label{eq88} \; \; .
\end{eqnarray}

\noindent
{\it The above expression is the second key result of the paper}. 
It has the form of the celebrated Arrhenius expression for rate coefficient
with the classical activation energy $E_b^c$ in the exponential factor 
and a complicated $\Lambda$ and $D$ dependent quantity in the pre-exponential 
factor. As noted earlier 
in the detailed discussion of quantum diffusion coefficient
in the context of Eq.(\ref{eq68}), the diffusion coefficient $D(E)$ is
contributed by the three factors and has to be evaluated at the barrier top.
The main effect of the pre-exponential factor is that the rate becomes 
proportional to the damping coefficient and the memory kernel results in
the decrease of pre-factor for increasing correlation time. The structure 
of the rate expression (\ref{eq88}) suggests that it has the same form
of the pre-exponential factor as that of H\"anggi and Weiss \cite{hw} 
although its content is quantum mechanical in character. 
The quantum mechanical content of the rate expression lies in several
quantities, e.g., quantum diffusion coefficient $D (E_b^c)$, quantum
analogue of $k_BT$, $\Lambda$ as given by (\ref{eq69}) and (\ref{eq65}),
respectively. The frequency at the bottom of the well, $\omega_0$ as well as
the classical activation energy $E_b^c$ get modified by quantum correction
$\Delta Q_0''$ and $\Delta Q_0$ terms. The result of H\"anggi and Weiss
\cite{hw} for the classical non-Markovian case can then be appropriately 
recovered. It is thus apparent that quantum correction terms in the
exponential factor is (\ref{eq88}) depends on the nature of the potential
which in turn determines the rate. In what follows in the next section
we illustrate the results with a specific cubic potential of Kramers' form.
We mention in passing that throughout the treatment the noise intensity 
needs to be small for the result (\ref{eq88}) to be a good description
of the activated process controlled by energy diffusion.


\section{An Example With Cubic Potential}

We consider a model cubic potential of the form
$V ( \hat{X} ) = -(1/3) A \hat{X}^3 + B \hat{X}^2$. $A$ and $B$ are two
constant parameters of the problem with $A>0$ and $B>0$.
Then by virtue of Eq.(\ref{eq13}) we have the $c$-number form of quantum 
potential

\begin{equation}
\label{eq89}
V_q (x) = - \frac{A}{3} x^3 + B x^2 - 
A \langle \delta \hat{X}^2 (0) \rangle x + {\rm constant}
\end{equation}

\noindent
so that the time independent Hamiltonian is given by,

\begin{equation}
\label{eq90}
H ( x,v ) =  \frac {v^2}{2} - \frac {A}{3} x^3 + B x^2 - C x = E
\end{equation}

\noindent
where $C = A \hbar / (2 \sqrt {2B}) $. We have
used $\langle \delta \hat{X}^2 (0) \rangle = \hbar/(2 \sqrt{2B})$, 
the minimum uncertainty, and ignored the constant part in $V_q (x)$.
Here $C$ refers to quantum contribution to classical potential due to which
the minimum, the metastable point corresponding to $V_q (x)$ shifts to
$x_0 = C/(2B)$ (with respect to the corresponding classical metastable
minimum at $x=0$).

\noindent
Linearizing the potential $V_q (x)$ around $x_0$ we obtain

\begin{equation}
\label{eq95}
V_q (x) = V_q (x_0) 
+ \left ( B - \frac { AC }{ 2B } \right ) ( x - x_0 )^2 \; \; .
\end{equation}

\noindent
We then calculate the action $ J $, the usual form of which is given by, 

\begin{equation}
\label{eq96}
J = 2 \int_{x_1}^{x_2} v dx 
\end{equation}

\noindent
$ x_1 $ and $ x_2 $ are the two turning points of oscillation for which
$ v $ is equal to zero and they jointly correspond to a particular value
of the system energy $ E $. In principle, they are the first two roots 
(in ascending order of magnitude) of the cubic equation 

\begin{equation}
\label{eq97}
\frac {A}{3} x^3 - B x^2 + C x + E = 0
\end{equation}

\noindent
the third root being irrelevant for the present purpose. $x_1$ and $x_2$
however can be approximately calculated by simply putting $V_q (x) = E$
(since $v=0$ is the turning point) in (\ref{eq95}) and solving the resulting 
equation for $x$.

\begin{equation}
\label{eq98}
x_{1,2} \simeq x_0 \mp \left ( \frac{ E - V_q (x_0) }
{ B } \right )^{1/2} \; \; .
\end{equation}

\noindent
Putting the value of $ v $ from Eq.(\ref{eq90}) in
Eq.(\ref{eq96}) we get the action integral in the form,

\begin{equation}
\label{eq99}
J = 2\sqrt{2} \int_{x_1}^{x_2} \left [ ( E - B x^2 ) + 
\left ( \frac {A}{3} x^3 + C x \right ) \right ]^{1/2} dx \; \; .
\end{equation}

\noindent
Putting Eq.(\ref{eq66}) in Eq.(\ref{eq69}) we can express the 
quantum diffusion co-efficient in terms of the action as,

\begin{equation}
\label{eq100}
\tilde D (J) = \omega (J) 2 \hbar \overline {\omega}_0 
\left [ \overline {n} ( \overline {\omega}_0 ) + \frac{1}{2} \right ] 
\sum_{n = 1}^\infty n^2 | x_n |^2
\int_0^\infty dt \beta (t) \cos [n \omega (J) t ]
\end{equation}

\noindent
where we have replaced $ D(E) $ by $ \tilde D(J) $ to emphasize the change
made in the argument. Furthermore for unit mass
of the Brownian particle we may write \cite{cn}

\begin{equation}
\label{eq101}
\omega (J) \sum _{n = -\infty}^\infty n^2 | x_n (J) |^2 = J
\end{equation}

\noindent
Putting Eq.(\ref{eq101}) in Eq.(\ref{eq100}) the diffusion coefficient
can be approximately expressed as

\begin{equation}
\label{eq102}
\tilde D (J) \simeq 2 J \hbar \overline {\omega}_0 
\left [ \overline{n} ( \overline{\omega}_0 ) + \frac{1}{2} \right ] 
\int_0^\infty dt \beta (t) \cos [n \omega (J) t ] \; \; .
\end{equation}

\noindent
For the present form of model potential we also have 
$\Delta Q (x) = - C x$ for which $\Delta Q_0 = 0$,
$\Delta Q_0'' = 0$ and 
$\Delta Q_0' = - C$. With these expressions for quantum contributions 
and making use of 
Eq.(\ref{eq102}) in Eq.(\ref{eq88}) we have the final expression for
the escape rate as,

\begin{equation}
k = \frac{\omega_0}{2\pi}
\left [ \frac { \{ 1 + (4 \alpha \Lambda^2) / \tilde{D} (J_b) \}^{1/2} - 1 }
{ \{ 1 + (4 \alpha \Lambda^2) / \tilde{D} (J_b) \}^{1/2} + 1 } \right ]
\frac { \tilde{D} (J_b) }{ \Lambda^2 } 
\exp \left [ -\frac {1}{\Lambda} 
\left \{ E_b^c + \frac{A^2 \hbar^2}{4 \omega_0^3}
\right \} \right ] \label{eq103}
\end{equation}

\noindent
Here $ J_b $ denotes the value of the action of the system at the barrier
top. It should be noted that it includes both the classical and quantum
contributions. $E_b^c$ corresponds to classical activation energy which
gets modified by a contribution due to quantum correction
entangled with the nonlinearity of the potential. It is important
to note that the positivity of the factor
$(A^2 \hbar^2)/(4 \omega_0^3 \Lambda)$ in the exponential in (\ref{eq103})
results in a larger effective activation barrier which causes a net reduction
of the full rate below its corresponding classical value. This is in good
agreement with the earlier observation by Griff {\it et al} \cite{griff}
and is somewhat counterintuitive - as emphasized by H\"anggi {\it et al}
\cite{htb} to the fact that full rate comprises classical rate plus
zero temperature tunneling. 
The quantum reduction of total rate in the weak friction
regime is a manifestation of interplay of thermal noise and quantum
fluctuation and is expected to be pronounced for systems with flat barriers,
commonly encountered in absorption-desorption processes in surface 
phenomena \cite{htb}.


\section{Conclusions}

Based on a true quantum phase space distribution function and an
ensemble average procedure we have derived a generalized Kramers'
equation for energy diffusion and analyzed the quantum transmission
coefficient associated with the rate coefficient within a full quantum 
mechanical framework in the low friction regime. The present formulation
is a complementary follow-up to our recent work \cite{dbbr} on quantum
Kramers' theory in the spatial diffusion limited regime. The main conclusions
of this study are the following:

(i) The proposed Kramers' equation in the energy diffusion regime is an
exact quantum analogue of non-Markovian classical Kramers' equation
derived by CN \cite{cn} in eighties. The equation retains
its full validity both in the classical and vacuum limits at arbitrary 
temperature and noise correlation of the heat bath.

(ii) The generalized quantum rate coefficient for the
decay from a metastable well reduces
to Kramers'-H\"anggi-Weiss rate \cite{hw} in the classical limit and to
pure weak dissipative tunneling rate in the quantum limit at zero
temperature.

(iii) While in the intermediate to strong damping regime the total Kramers'
rate comprising classical as well as quantum rate
is always higher than the corresponding classical rate, the notable feature
in the weak friction regime $($ for a metastable potential $)$ 
is a net quantum reduction of the total rate below
its corresponding classical value. This is in conformity with the earlier
observation in this context \cite{griff}.

(iv) While the existing methods of calculation of quantum Kramers' rate
is based on path integral techniques \cite{htb,garg,ingold,uw,fv,hibbs}, 
we rely on a canonical quantization
procedure and true probability distribution function of $c$-number variables.
To the best of our knowledge the implementation of a differential equation
and its solution as a boundary value problem have not been tried up to date
for quantum Kramers' problem. The methodology as pursued here allows us
to apply classical techniques for the quantum problem of barrier crossing
dynamics.

(v) The quantum effects appear in the present formulation in two different 
ways. The nonlinear part of the potential of the system gives rise to
quantum dispersion, while the heat bath imparts quantum noise. An 
important advantage of the present method is that it is possible to
incorporate quantum corrections to all orders and one need not invoke any
semiclassical technique which is almost always used in the practical
evaluation of the formal functional integrals.

The present scheme of mapping of the quantum theory of Brownian motion
in energy space into a classical form offers an opportunity to generate
quantum noise as classical $c$-numbers and study numerically the
quantum stochastic dynamics independent of path integral Monte Carlo
techniques \cite{topaler,liao,Creswick}. 
We hope to address this issue in a future communication.

\acknowledgments
The authors are indebted to the Council of Scientific and Industrial Research
(CSIR), Government of India, for financial support under grant No.
01/(1740)/02/EMR-II.


\begin{appendix}


\section{Calculation of moments}

Some details of the calculations involving the iterations (upto the third
order) for determination of the moments have been shown here. The procedure
followed here is that of classical theories of Carmeli and Nitzan \cite{cn}.
We have stressed the steps for which the quantum contributions form
essentially new content.

\subsection{First Iteration}
               
Inserting Eq.(\ref{eq50}) and Eq.(\ref{eq51}) in Eq.(\ref{eq48}) we have,

\begin{eqnarray}
\Delta & J_t^{(1)} & ( \tau )  =  
-\sum_{n = -\infty}^\infty \sum_{m = -\infty}^\infty B_{nm} 
\exp [i(n + m) \phi] \int_0^\tau ds \exp [i(n + m) \omega s] \nonumber \\
& + & \sum_{n = -\infty}^\infty 
\sigma_n \exp (in \phi) \int_0^\tau ds F (s) \exp (in \omega s) +
\sum_{n = -\infty}^\infty 
\sigma_n \exp (in \phi) \int_0^\tau ds g (s) \exp (in \omega s)
\label{eqa1}
\end{eqnarray}

\noindent
where we have suppressed the arguments of quantities $ B_{nm}$, $\sigma_n $
and $ \phi $ for the sake of brevity. Now, from Eq.(\ref{eq17})
and the argument following Eq.(\ref{eq37}) we can infer that 
$ \omega \tau \gg 1 $. So we can write

\begin{eqnarray}
\label{eqa2}
\int_0^\tau ds \exp [i(n + m) \omega s] & \simeq & \tau \delta_{n,-m}  \\
{\rm for \; which} \; \; \; \; \;
\sum_{n,m = -\infty}^\infty B_{nm} \exp [i(n + m)] \tau \delta_{n,-m}
& = & \tau \sum_{n = -\infty}^\infty B_{n,-n}
\label{eqa3}
\end{eqnarray}

\noindent
Thus,

\begin{eqnarray}
\Delta J_t^{(1)} (\tau) =
& - \tau & \sum_{n = -\infty}^\infty B_{n,-n} 
+ \sum_{n = -\infty}^\infty 
\sigma_n \exp (in \phi) \int_0^\tau ds F (s) \exp (in \omega s)  \nonumber \\
& + & \sum_{n = -\infty}^\infty 
\sigma_n \exp (in \phi) \int_0^\tau ds g (s) \exp (in \omega s)
\label{eqa4}
\end{eqnarray}

Similarly from Eq.(\ref{eq49}) we have 

\begin{eqnarray}
\Delta \phi_t^{(1)} (\tau) =  & \omega & \tau + 
\tau \sum_{n = -\infty}^\infty C_{n,-n}  
- \sum_{n = -\infty}^\infty 
\mu_n \exp (in \phi) \int_0^\tau ds F (s) \exp (in \omega s)  \nonumber \\
& - & \sum_{n = -\infty}^\infty 
\mu_n \exp (in \phi) \int_0^\tau ds g (s) \exp (in \omega s)
\label{eqa5}
\end{eqnarray}

\noindent
Along with these we will also require the statistical averages of the above
expressions for calculation of the moments. Thus,

\begin{equation}
\label{eqa6}
\langle \Delta J_t^{(1)} (\tau) \rangle_S = 
- \tau \sum_{n = -\infty}^\infty B_{n,-n} 
+ \sum_{n = -\infty}^\infty 
\sigma_n \exp (in \phi) \int_0^\tau ds g (s) \exp (in \omega s)
\end{equation}

\noindent
where we have used (\ref{eqn10}).
We can proceed further with Eq.(\ref{eqa6}) 
and cast it in a more transparent form. In this context it is worth 
mentioning  that so far as the quantum noise $ g(t) $ is concerned, 
we take only the significant terms from Eq.(\ref{eq11}) which result in
terms linear in $ \tau $. Referring back to Eq.(\ref{eq44}) we then have,

\begin{eqnarray}
-\tau \sum_{n = -\infty}^\infty B_{n,-n} & = &
-\tau \sum_{n = -\infty}^\infty n^2 \omega | x_n |^2 \tilde{\beta}_n (\omega)
\nonumber  \\  
& = & -2\tau \sum_{n = 1}^\infty n^2 \omega | x_n |^2 \tilde{\beta}_n^c (\omega) 
\label{eqa7}
\end{eqnarray}

\noindent
using Eq.(\ref{eq59a}). The second expression of Eq.(\ref{eqa6}) has been 
shown to be negligible in Appendix-B. Thus

\begin{equation}
\label{eqa8}
\langle \Delta J_t^{(1)} (\tau) \rangle_S =
-2\tau \sum_{n = 1}^\infty n^2 \omega | x_n |^2 \tilde{\beta}_n^c (\omega) 
\end{equation}

\noindent
Similarly from Eq.(\ref{eqa5}) we have,

\begin{equation}
\label{eqa9}
\langle \Delta \phi_t^{(1)} (\tau) \rangle_S =
\omega \tau + \tau \sum_{n = 1}^\infty n \omega \frac { d |x_n|^2 }{ dJ }
\tilde {\beta}_n^s - \tau f_0^\prime \mu_0 t_c
\end{equation}

\noindent
In deriving Eq.(\ref{eqa9}) we have used Eq.(\ref{eq43}) and (\ref{eq45}).
Otherwise, the way leading to Eq.(\ref{eqa9}) is similar to that of
Eq.(\ref{eqa8}). The origin of the last term is the same integral that 
appears in Eq.(\ref{eqa6}) and is shown in Appendix-B.

\subsection{Second Iteration}

\noindent
Here we insert Eq.(\ref{eqa4}) and (\ref{eqa5}) into the right hand side's of 
Eq.(\ref{eq48}) and Eq.(\ref{eq49}). Keeping in mind Eq.(\ref{eq50}) and  
(\ref{eq51}) we expand the functions of $ \Delta J_t (s) $ and
$ \Delta \phi_t (s) $ , viz. $ B_{nm},\; C_{nm},\; \sigma_n $ and $ \mu_n $,
keeping terms upto the first order only. Thus,

\begin{mathletters}

\begin{eqnarray}
\label{eqa10a}
B_{nm} [ J(t) + \Delta J_t (s) ] & = & B_{nm} + B_{nm}^\prime 
\Delta J_t (s) \\   {\rm with} \; \; \; \;
\label{eqa10b}
B_{nm} & = & B_{nm} [ J(t) ]  \\  {\rm and} \; \; \; \;
\label{eqa10c}
B_{nm}^\prime & = & \left [ \frac { dB_{nm} }{ dJ } \right ]_{J = J(t)}
\end{eqnarray}

\end{mathletters}

\noindent
The expansion has been done about $ \Delta J_t^{(0)} (s) = 0 $ and
$ \Delta \phi_t^{(0)} (s) = \omega s $. The same expansion procedure is 
followed for $ C_{nm},\; \sigma_n,\; \mu_n $ and the exponentials as well, 
occurring in Eq.(\ref{eq48}) and (\ref{eq49}). We therefore obtain,

\begin{eqnarray}
\Delta & J_t^{(2)} & (\tau) = \Delta J_t^{(1)} (\tau) - 
\sum_{n = -\infty}^\infty \sum_{m = -\infty}^\infty \sum_{l = -\infty}^\infty
[ B_{nm}^\prime \sigma_l - i(n + m) B_{nm} \mu_l ] \exp [ i(n + m + l) \phi ]
\nonumber \\  & \times &
\int_0^\tau ds \int_0^s ds_1 F (s_1) \exp [ i(n + m)\omega s + il\omega s_1 ]
- \sum_{n = -\infty}^\infty \sum_{m = -\infty}^\infty \sum_{l=-\infty}^\infty
[ B_{nm}^\prime \sigma_l - \nonumber \\   
& i & (n + m) B_{nm} \mu_l ] \exp [ i(n + m + l) \phi ]
\int_0^\tau ds \int_0^s ds_1 g (s_1) \exp [ i(n + m)\omega s + il\omega s_1 ]
\nonumber  \\  
& + & \sum_{n = -\infty}^\infty \sum_{l = -\infty}^\infty
( \sigma_n^\prime \sigma_l - in \sigma_n \mu_l ) \exp [ i(n + l) \phi ]
\int_0^\tau ds \int_0^s ds_1 F (s) F (s_1) \exp (in\omega s + il\omega s_1)
\nonumber \\
& + & \sum_{n = -\infty}^\infty \sum_{l = -\infty}^\infty
( \sigma_n^\prime \sigma_l - in \sigma_n \mu_l ) \exp [ i(n + l) \phi ]
\int_0^\tau ds \int_0^s ds_1 F (s) g (s_1) \exp (in\omega s + il\omega s_1)
\nonumber \\
& + & \sum_{n = -\infty}^\infty \sum_{l = -\infty}^\infty
( \sigma_n^\prime \sigma_l - in \sigma_n \mu_l ) \exp [ i(n + l) \phi ]
\int_0^\tau ds \int_0^s ds_1 g (s) F (s_1) \exp (in\omega s + il\omega s_1)
\nonumber \\
& + & \sum_{n = -\infty}^\infty \sum_{l = -\infty}^\infty
( \sigma_n^\prime \sigma_l - in \sigma_n \mu_l ) \exp [ i(n + l) \phi ]
\int_0^\tau ds \int_0^s ds_1 g (s) g (s_1) \exp (in\omega s + il\omega s_1)
\nonumber \\ 
\label{eqa11}
\end{eqnarray}

\noindent
The derivation this form of Eq.(\ref{eqa11}) requires two important steps to 
be followed. The first step 
is just a Taylor expansion. While putting Eq.(\ref{eqa4}) and (\ref{eqa5})
in Eq.(\ref{eq48}) and  Eq.(\ref{eq49}), we come across several integrals
which contain one quantity in common in the integrands. This is an 
expression of the form $ \exp [ ik\{\phi (t) + \Delta \phi_t^{(1)}(s)\} ] $,
where $ k $ is an integer. The first part, i.e. $ \exp [ ik \phi (t) ]  $,
being a function of $ t $, can be taken outside the integral while the second 
part can be dealt with as follows: 

\begin{eqnarray}
\exp [ ik \{ \Delta \phi_t^{(1)}(s) \} ] & = &
\exp [ ik \{ \Delta \phi_t^{(1)}(s) - \Delta \phi_t^{(0)}(s) \} ]
\exp [ ik \{ \Delta \phi_t^{(0)}(s) \} ]   \nonumber   \\
& = &  [ 1 + ik \{ \Delta \phi_t^{(1)}(s) - \omega s \} ] \exp( ik \omega s )
\label{eqa12}
\end{eqnarray}

\noindent
Here we have used Eq.(\ref{eq51}) and as an essential step (second) 
discarded the nonlinear terms. Also, since 
$ \Delta J_t^{(1)}(\tau) $ and $ \Delta \phi_t^{(1)}(\tau) $ are of order
$ O (\tau) $, we have

\begin{equation}
\label{eqa13}
\Delta J_t^{(1)} (s) \exp [ ik \Delta \phi_t^{(1)}(s) ]
\simeq \Delta J_t^{(1)} (s) \exp( ik \omega s )
\end{equation}

\noindent
Guided by the same physical reasoning, we assert that all integrals of the
form $ \int_0^\tau ds \; s F (s) $, with $ F(s) $ being finite as 
$ s \rightarrow 0 $, yield terms of the order $ \tau^n \; (n>1) $, and are  
hence neglected. In a similar way, we can obtain the second iteration on
$ \phi $. Thus,

\begin{eqnarray}
\Delta & \phi_t^{(2)} & (\tau) = \Delta \phi_t^{(1)} (\tau) + 
\omega^\prime \sum_{n = -\infty}^\infty \sigma_n \exp (in \phi)
\int_0^\tau ds \int_0^s ds_1 F (s_1) \exp (in \omega s_1) 
\nonumber  \\  & + &
\omega^\prime \sum_{n = -\infty}^\infty \sigma_n \exp (in \phi)
\int_0^\tau ds \int_0^s ds_1 g (s_1) \exp (in \omega s_1) +
\sum_{n = -\infty}^\infty \sum_{m = -\infty}^\infty \sum_{l = -\infty}^\infty
[ C_{nm}^\prime \sigma_l -  
\nonumber  \\  & i &(n + m)
C_{nm} \mu_l ] \exp [ i(n + m + l) \phi ]
\int_0^\tau ds \int_0^s ds_1 F (s_1) \exp [ i(n + m)\omega s + il\omega s_1 ]
\nonumber  \\  & + &
\sum_{n = -\infty}^\infty \sum_{m = -\infty}^\infty \sum_{l = -\infty}^\infty
[ C_{nm}^\prime \sigma_l -  
i (n + m) C_{nm} \mu_l ] \exp [ i(n + m + l) \phi ]
\int_0^\tau ds \int_0^s ds_1 g (s_1) 
\nonumber  \\  & \; & \; \; \;  
\times \exp [ i(n + m)\omega s + il\omega s_1 ]
\nonumber  \\  & - &
\sum_{n = -\infty}^\infty \sum_{l = -\infty}^\infty
( \mu_n^\prime \sigma_l - in \mu_n \mu_l ) \exp [ i(n + l) \phi ]
\int_0^\tau ds \int_0^s ds_1 F (s) F (s_1) \exp (in\omega s + il\omega s_1)
\nonumber \\   & - &
\sum_{n = -\infty}^\infty \sum_{l = -\infty}^\infty
( \mu_n^\prime \sigma_l - in \mu_n \mu_l ) \exp [ i(n + l) \phi ]
\int_0^\tau ds \int_0^s ds_1 F (s) g (s_1) \exp (in\omega s + il\omega s_1)
\nonumber \\   & - &
\sum_{n = -\infty}^\infty \sum_{l = -\infty}^\infty
( \mu_n^\prime \sigma_l - in \mu_n \mu_l ) \exp [ i(n + l) \phi ]
\int_0^\tau ds \int_0^s ds_1 g (s) F (s_1) \exp (in\omega s + il\omega s_1)
\nonumber \\   & - &
\sum_{n = -\infty}^\infty \sum_{l = -\infty}^\infty
( \mu_n^\prime \sigma_l - in \mu_n \mu_l ) \exp [ i(n + l) \phi ]
\int_0^\tau ds \int_0^s ds_1 g (s) g (s_1) \exp (in\omega s + il\omega s_1)
\nonumber \\ 
\label{eqa14}
\end{eqnarray}

\noindent
We next require to calculate the averages
$ \langle \Delta J_t^{(2)} (\tau) \rangle_S $ and
$ \langle \Delta \phi_t^{(2)} (\tau) \rangle_S $. 
By virtue of Eq.(\ref{eqn10}),
the former yields,

\begin{eqnarray}
\langle \Delta & J_t^{(2)} & (\tau) \rangle_S =
\langle \Delta J_t^{(1)} (\tau) \rangle_S
- \sum_{n = -\infty}^\infty \sum_{m = -\infty}^\infty \sum_{l=-\infty}^\infty
[ B_{nm}^\prime \sigma_l - i(n + m) B_{nm} \mu_l ] 
\exp [ i(n + m + l) \phi ] 
\nonumber \\   & \times &
\int_0^\tau ds \int_0^s ds_1 g (s_1) \exp [ i(n + m)\omega s + il\omega s_1 ]
\nonumber \\   & + & 
\sum_{n = -\infty}^\infty \sum_{l = -\infty}^\infty
( \sigma_n^\prime \sigma_l - in \sigma_n \mu_l ) \exp [ i(n + l) \phi ]
\int_0^\tau ds \int_0^s ds_1 \langle F (s) F (s_1) \rangle
\exp (in\omega s + il\omega s_1)
\nonumber \\   & + & 
\sum_{n = -\infty}^\infty \sum_{l = -\infty}^\infty
( \sigma_n^\prime \sigma_l - in \sigma_n \mu_l ) \exp [ i(n + l) \phi ]
\int_0^\tau ds \int_0^s ds_1 g (s) g (s_1) 
\exp (in\omega s + il\omega s_1)
\nonumber \\ 
\label{eqa15}
\end{eqnarray}

\noindent
In Eq.(\ref{eqa15}) we encounter three double integrals. 
The first and consequently
the third are shown to be negligible in Appendix-B. For the second integral,
viz. $ \int_0^\tau ds \int_0^s ds_1 \langle F (s) F (s_1) \rangle_S
\exp (in\omega s + il\omega s_1) $, or
$ \int_0^\tau ds \int_0^s ds_1 C (s - s_1)\exp (in\omega s + il\omega s_1) $,
we refer to Ref.~\cite{cn}. In Appendix-C of Ref.~\cite{cn} 
this type of integral of more general form have been evaluated. 
Here we state the results only.

\begin{eqnarray}
I_{n,l}^{(j)} & = &
\int_0^\tau ds \int_0^s ds_1 \int_0^{s_1} ds_2 \; ... \int_0^{s_j - 1} ds_j
C(s - s_j) \exp [ in \omega s + il \omega s_j ]
\nonumber \\  & \simeq &
\frac {1}{ (j - 1)! }\left ( \frac {i}{l} \right )^{(j-1)} 
\frac { d^{j - 1} \tilde{\beta}_l (\omega) }{ d \omega^ {j - 1} } 
\tau \delta_{n,-l}
\label{eqa16}
\end{eqnarray}

\noindent
The important physical consideration here is to neglect the terms of order 
$ \tilde{C} (\omega) / \omega $.
The rest of the task amounts to solving the above
integral, also known as the Dirichlet's condition for multiple integrals.
Thus the integrals of the third term in the right hand side of 
Eq.(\ref{eqa15}) reduce to,

\begin{equation}
\label{eqa17}
\int_0^\tau ds \int_0^s ds_1 C (s - s_1) \exp (in\omega s + il\omega s_1) =
\tau \tilde{\beta}_l (\omega) \delta_{n,-l}
\end{equation}

\noindent
Putting Eq.(\ref{eqa17}) back into Eq.(\ref{eqa15}) yields, after a little 
algebra,

\begin{eqnarray*}
\langle \Delta J_t^{(2)} (\tau) \rangle_S = 
\langle \Delta J_t^{(1)} (\tau) \rangle_S + 2\tau \sum_{n = 1}^\infty
n^2 \frac { d | x_n |^2 }{ dJ } \tilde{C}_n^c (\omega) 
\end{eqnarray*}

\noindent
which, after using Eq.(\ref{eqa8}) reduces to,

\begin{equation}
\label{eqa18}
\langle \Delta J_t^{(2)} (\tau) \rangle_S = 
-2\tau \sum_{n = 1}^\infty n^2 
\left [ \omega | x_n |^2 \tilde{\beta}_n^c (\omega)
- \frac { d | x_n |^2 }{ dJ } \tilde{C}_n^c (\omega) \right ]
\end{equation}

\noindent
In a similar manner we also get,

\begin{equation}
\label{eqa19}
\langle \Delta \phi_t^{(2)} (\tau) \rangle_S = \omega \tau +
\tau \sum_{n = 1}^\infty n \left [ \omega \tilde{\beta}_n^s 
- \tilde{C}_n^s \frac{d}{dJ} \right ]
\frac { d | x_n |^2 }{ dJ } - \tau f_0^\prime \mu_0 t_c
\end{equation}

\noindent
Next we step out for the final iteration stage, the third one.

\subsection{Third Iteration}

Inserting Eq.(\ref{eqa11}) and Eq.(\ref{eqa14}) in Eq.(\ref{eq48}) and
Eq.(\ref{eq49}) as before we get the expansions for $ \Delta J_t^{(3)} 
(\tau) $ and $ \Delta \phi_t^{(3)} (\tau) $which we write in the following
convenient form.

\begin{eqnarray}
\Delta J_t^{(3)} (\tau) & = &
\Delta J_t^{(2)} (\tau) 
- \sum_{n = -\infty}^\infty \sum_{m = -\infty}^\infty B_{nm}^\prime
\exp [i(n + m) \phi] \int_0^\tau ds [ \Delta J_t^{(2)} (s) -
\Delta J_t^{(1)} (s) ] 
\nonumber  \\
& & \times \exp [i(n + m) \omega s]  
\nonumber  \\
& & - \sum_{n = -\infty}^\infty \sum_{m = -\infty}^\infty i(n + m) B_{nm}
\exp [i(n + m) \phi] \int_0^\tau ds [ \Delta \phi_t^{(2)} (s) -
\Delta \phi_t^{(1)} (s) ] 
\nonumber  \\
& & \times \exp [i(n + m) \omega s]  
\nonumber  \\
& &  + \sum_{n = -\infty}^\infty \sigma_n^\prime
\exp (in \phi) \int_0^\tau ds [ \Delta J_t^{(2)} (s) -
\Delta J_t^{(1)} (s) ] F (s) \exp (in \omega s)  
\nonumber  \\
& & + \sum_{n = -\infty}^\infty in \sigma_n
\exp (in \phi) \int_0^\tau ds [ \Delta \phi_t^{(2)} (s) -
\Delta \phi_t^{(1)} (s) ] F (s) \exp [in \omega s]  
\nonumber  \\
& & + \sum_{n = -\infty}^\infty \sigma_n^\prime
\exp (in \phi) \int_0^\tau ds [ \Delta J_t^{(2)} (s) -
\Delta J_t^{(1)} (s) ] g (s) \exp (in \omega s)  
\nonumber  \\
& & + \sum_{n = -\infty}^\infty in \sigma_n
\exp (in \phi) \int_0^\tau ds [ \Delta \phi_t^{(2)} (s) -
\Delta \phi_t^{(1)} (s) ] g (s) \exp [in \omega s]  
\label{eqa20}
\end{eqnarray}

\noindent
and

\begin{eqnarray}
\Delta \phi_t^{(3)} (\tau) & = &
\Delta \phi_t^{(2)} (\tau) 
+ \omega^\prime \int_0^\tau ds [ \Delta J_t^{(2)}(s) - \Delta J_t^{(1)}(s) ]
\nonumber  \\
& &  + \sum_{n = -\infty}^\infty \sum_{m = -\infty}^\infty C_{nm}^\prime
\exp [i(n + m) \phi] \int_0^\tau ds [ \Delta J_t^{(2)} (s) -
\Delta J_t^{(1)} (s) ] \exp [i(n + m) \omega s]  
\nonumber  \\
& & +  \sum_{n = -\infty}^\infty \sum_{m = -\infty}^\infty i(n + m) C_{nm}
\exp [i(n + m) \phi] \int_0^\tau ds [ \Delta \phi_t^{(2)} (s) -
\Delta \phi_t^{(1)} (s) ] 
\nonumber  \\  
& & \times \exp [i(n + m) \omega s]  
\nonumber  \\
& & - \sum_{n = -\infty}^\infty \mu_n^\prime
\exp (in \phi) \int_0^\tau ds [ \Delta J_t^{(2)} (s) -
\Delta J_t^{(1)} (s) ] F (s) \exp (in \omega s)  
\nonumber  \\
& & - \sum_{n = -\infty}^\infty in \mu_n
\exp (in \phi) \int_0^\tau ds [ \Delta \phi_t^{(2)} (s) -
\Delta \phi_t^{(1)} (s) ] F (s) \exp (in \omega s)  
\nonumber  \\
& & - \sum_{n = -\infty}^\infty \mu_n^\prime
\exp (in \phi) \int_0^\tau ds [ \Delta J_t^{(2)} (s) -
\Delta J_t^{(1)} (s) ] g (s) \exp (in \omega s)  
\nonumber  \\
& & +\sum_{n = -\infty}^\infty in \mu_n
\exp (in \phi) \int_0^\tau ds [ \Delta \phi_t^{(2)} (s) -
\Delta \phi_t^{(1)} (s) ] g (s) \exp ( in \omega s )  
\label{eqa21}
\end{eqnarray}

\noindent
It is needless to carry out further iterations because they yield terms of
order $ \tilde{C} (\omega) / \omega $ even in the order of $ \tau $,
and hence are negligible. From this point we therefore proceed to calculate 
the moments $ \langle ( \Delta J )^m ( \Delta \phi )^k \rangle_S $ as the 
right hand side of Eq.(\ref{eq37}) demands. While performing the averaging, some 
integrals occur in common with both $ \langle \Delta J_t^{(3)}(\tau) 
\rangle_S $ and $ \langle \Delta\phi_t^{(3)} (\tau) \rangle_S $. 
They are as follows:-

\begin{mathletters}

\begin{eqnarray}
\label{eqa22a}
I_1 & = & \int_0^\tau ds \langle [ \Delta J_t^{(2)} (s) -  
\Delta J_t^{(1)} (s) ] \rangle_S  \\
\label{eqa22b}
I_2 & = & \int_0^\tau ds \langle [ \Delta J_t^{(2)} (s) -  
\Delta J_t^{(1)} (s) ] \rangle_S \exp [ i(n + m) \omega s ]   \\
\label{eqa22c}
I_3 & = & \int_0^\tau ds \langle [ \Delta \phi_t^{(2)} (s) -  
\Delta \phi_t^{(1)} (s) ] \rangle_S \exp [ i(n + m) \omega s ]  \\
\label{eqa22d}
I_4 & = & \int_0^\tau ds \langle [ \Delta J_t^{(2)} (s) -  
\Delta J_t^{(1)} (s) ] F (s) \rangle_S \exp ( in \omega s )    \\
\label{eqa22e}
I_5 & = & \int_0^\tau ds \langle [ \Delta \phi_t^{(2)} (s) -  
\Delta \phi_t^{(1)} (s) ] F (s) \rangle_S \exp ( in \omega s )  \\
\label{eqa22f}
I_6 & = & \int_0^\tau ds \langle [ \Delta J_t^{(2)} (s) -  
\Delta J_t^{(1)} (s) ] \rangle_S g (s)  \exp ( in \omega s )   \\
\label{eqa22g}
I_7 & = & \int_0^\tau ds \langle [ \Delta \phi_t^{(2)} (s) -  
\Delta \phi_t^{(1)} (s) ] \rangle_S g (s) \exp ( in \omega s )   
\end{eqnarray}

\end{mathletters}

\noindent
Integrals $ I_2 $ to $ I_7 $ are in common with 
$ \langle \Delta J_t^{(3)} (\tau) \rangle_S $ and
$ \langle \Delta \phi_t^{(3)} (\tau) \rangle_S $ while $ I_1 $ occurs in the
latter only. The integrands containing $ F (.) $ alone or in a product with
$ g (.) $, on averaging gives zero by virtue of Eq.(\ref{eqn10}). And the
integrands containing $ g (.) $ alone or two $ g $'s in product, amounts to
negligible contributions (see Appendix-B). The integrands containing the 
correlation average of two $ F $'s and not obeying (or can not be recast) 
the form of Eq.(\ref{eqa16}) have all been shown to be negligible by
CN in their Appendix-C \cite{cn}. Such integrals as stated above
constitute the bodies of $ I_1 $ to $ I_3 $. For $ I_4 $ through $ I_7 $,
some integrals contain product of three terms viz.
$ \langle F(.) F(.) F(.) \rangle_S$, $\langle F(.) \rangle_S g(.) g(.)$,
$\langle F(.) F(.) \rangle_S g(.)$, $ g(.) g(.) g(.) $ etc.
The first two type of integrals vanish, 
the former due to Eq.(\ref{eqn10}) and the
latter due to the Gaussian property of $ F $. The last two types vanish,
the latter following from Appendix-B and the former has been shown to be
negligible in Appendix-C. The only non-zero contribution comes from
$ I_5 $, the integral as a whole appears as
$ \omega^\prime \sum_{l = -\infty}^\infty \sigma_l \exp (il \phi)
\int_0^\tau ds \int_0^s ds_1 \int_0^{s_1} ds_2 C(s - s_2)
\exp ( in \omega s + il \omega s_2 ) $, where we have used Eq.(\ref{fdr}).
Using the result of Eq.(\ref{eqa16}) followed by a little bit of algebra
we eventually arrive at the value of $ I_5 $. Thus,

\begin{eqnarray}
\label{eqa23}
I_5 & = & - \frac {i \tau \omega^\prime}{n} \sigma_{-n} \exp ( -in \phi )
\frac { d\tilde{C}_{-n} (\omega) }{ d\omega }   \\  {\rm and } \; \; \; \; \;
\label{eqa24}
I_1 & = & I_2 = I_3 = I_4 = I_6 = I_7 = 0
\end{eqnarray}

\noindent
Thus we have,

\begin{equation}
\label{eqa25}
\langle \Delta J_t^{(3)} (\tau) \rangle_S = 
\langle \Delta J_t^{(2)} (\tau) \rangle_S + \sum_{n = -\infty}^\infty
in \sigma_n \exp (in \phi) [I_5]
\end{equation}

\noindent
which, by changing the dummy index from $ -n $ to $ n $ for this 
symmetric summation $ ( -\infty \; {\rm to} \;  +\infty ) $, and employing
Eq.(\ref{eq23}), (\ref{eq42}) and (\ref{eq59g}) and finally putting the value of
$ \langle \Delta J_t^{(2)} (\tau) \rangle_S $ from Eq.(\ref{eqa18}) we reach at,

\begin{equation}
\label{eqa26}
\langle \Delta J_t^{(3)} (\tau) \rangle_S = 
- 2 \tau \sum_{n = 1}^\infty n^2 \left [ \omega |x_n|^2 \tilde{\beta}_n^c (\omega)
- \frac {d}{dJ} \{ |x_n|^2 \tilde{C}_n^c (\omega) \} \right ]
\end{equation}

\noindent
Similar calculations establish the result,

\begin{equation}
\label{eqa27}
\langle \Delta \phi_t^{(3)} (\tau) \rangle_S = 
\omega \tau + \tau \sum_{n = 1}^\infty n \left [ \omega \tilde{\beta}_n^s
\frac{ d|x_n|^2 }{ dJ } - \frac{d}{dJ} \left ( \tilde{C}_n^s 
\frac { d|x_n|^2 }{ dJ } \right ) \right ] - \tau f_0^\prime \mu_0 t_c
\end{equation}

\noindent
For the calculations of the second moments
$ \langle ( \Delta J_t )^2 \rangle_S $, 
$ \langle ( \Delta \phi_t )^2 \rangle_S $
and $ \langle \Delta J_t \Delta \phi_t \rangle_S $ we refer the reader 
to the Appendices B and C of Ref.~\cite{cn}, where it has been clearly
shown that terms which do not appear in  $ \Delta J_t^{(1)} (\tau) $ and
$ \Delta \phi_t^{(1)} (\tau) $ lead to cross terms or square terms containing
three or more integrals over one or two $ C $ functions. All of them contain
higher powers of $ \tau $ or terms of order 
$ [ \tilde{C} (\omega) / \omega ]^N $ with $ N \ge 1 $, and are hence 
discarded. Thus we obtain the second moments as,

\begin{eqnarray}
\label{eqa28}
\langle ( \Delta J_t )^2 \rangle_S & = & 4 \tau \sum_{n = 1}^\infty n^2 |x_n|^2
\tilde{C}_n^c (\omega)      \\
\label{eqa29}
\langle \Delta J_t \Delta \phi_t \rangle_S & = & 0       \\
\label{eqa30}
\langle ( \Delta \phi_t )^2 \rangle_S & = & 4 \tau \sum_{n = 1}^\infty
\left |\frac {dx_n}{dJ} \right |^2 \tilde{C}_n^c (\omega)
\end{eqnarray}

\noindent
In obtaining the results of Eq.(\ref{eqa28}) to (\ref{eqa30}) we have used
Eq.(\ref{eqa10a}) to Eq.(\ref{eqa10c}) and the like, along with the fact
that higher order terms $ \tau^n $ and $ [ \tilde{C} (\omega) / \omega ]^n $
(with $ n>1 $) are negligibly small.


\section{Evaluation of integrals involving quantum corrections}

Here we calculate the terms involving single or multiple integrals of the 
quantum fluctuation $ g(.) $ keeping in mind the time scale of energy
diffusion. First of all we recall the structure of $ g (t) $ 
from Eq.(\ref{eq11}) and then refer to Eq.(\ref{eqa6}).
On the right hand side of that equation we have an expression of the form,
$ \sum_{n = -\infty}^\infty \sigma_n \exp ( in \phi ) \int_0^\tau ds g (s)
\exp ( in \omega s ) $. For brevity, we set

\begin{equation}
\label{eqb1}
\frac{1}{ (n - 1)! m!} \frac{ \partial ^m }{ \partial t^m } \left [ V^n 
\left ( x(t) \right ) \left \{ \langle \delta {\hat X} ^{(n-1)} (t) 
\rangle - \langle \delta {\hat X} ^{(n-1)} (0) \rangle \right \} \right ]
_{t=0} = \Upsilon_{mn} (0) 
\end{equation}

\noindent
so that, Eq.(\ref{eq11}) appears as

\begin{equation}
\label{eqb2}
g (t) = - \sum_{m = 0}^\infty \sum_{n = 3}^\infty t^m \Upsilon_{mn} (0) 
\end{equation}

\noindent
and we can write

\begin{eqnarray}
\sum_{n = -\infty}^\infty \sigma_n \exp ( in \phi ) \int_0^\tau ds g (s)
\exp ( in \omega s ) & = & 
\sum_{n = -\infty}^\infty \sum_{l = 3}^\infty
\sigma_n \exp ( in \phi ) 
\nonumber  \\
& & \times \left [ \sum_{k = 0}^\infty \Upsilon_{kl} (0)
\int_0^\tau ds \; s^k \exp ( in \omega s ) \right ]
\label{eqb3}
\end{eqnarray}

\noindent
From Eq.(\ref{eqb1}) it is obvious that

\begin{equation}
\label{eqb4}
\Upsilon_{0n} (0) = 0
\end{equation}

\noindent
and thus it follows that the right hand side of Eq.(\ref{eqb3}) 
assumes the form $ \sum_{n = -\infty}^\infty \sum_{l = 3}^\infty
\sigma_n \exp ( in \phi ) \sum_{k = 1}^\infty \Upsilon_{kl} (0)
\int_0^\tau ds s^k \exp ( in \omega s ) $ where the summation over the 
index $ k $ now extends from one to infinity. Within the integral the
exponential part fluctuates rapidly ( $ \omega $ being very large ), remaining
finite in the limit $ s \rightarrow 0 $. As it stands even for $ k = 2 $
such integrals yield terms of order $ \tau^p $ ( where $ p > 1 $ ) \cite{cn}. 
Thus, integrals of the above form are  
negligibly small and are hence discarded. The case of $ k = 1 $ calls for
a special attention. The above expression takes a more simple form. Thus,

\begin{eqnarray}
\sum_{n = -\infty}^\infty \sum_{l = 3}^\infty
\sigma_n \exp ( in \phi ) \sum_{k = 1}^\infty \Upsilon_{kl} (0)
\int_0^\tau ds s^k \exp ( in \omega s ) 
& \simeq &
\sum_{n = -\infty}^\infty \sum_{l = 3}^\infty
\sigma_n \exp ( in \phi ) \Upsilon_{1l} (0) 
\nonumber  \\
& & \times \int_0^\tau 
ds s \exp ( in \omega s ) 
\label{eqb5}
\end{eqnarray}

\noindent
In comparison to the rapidly varying exponential part, the linear part 
arising from the quantum fluctuation does not alter significantly within the
range of integration and hence the right hand side of Eq.(\ref{eqb5}) can be written as

\begin{eqnarray}
{\rm [ right \; hand \; side \; of \; Eq.(\ref{eqb5}) ] }
& = & \sum_{n = -\infty}^\infty \sum_{l = 3}^\infty
\sigma_n \exp ( in \phi ) \Upsilon_{1l} (0) t_c \tau \delta_{n,0}   
\nonumber  \\
& = & \sum_{n = -\infty}^\infty \sum_{l = 3}^\infty
(in x_n) \exp ( in \phi ) \Upsilon_{1l} (0) t_c \tau \delta_{n,0}
\; \; \; \; \; \; \; { [ \rm using \; Eq.(\ref{eq42}) ] } 
\nonumber  \\
& = & 0
\label{eqb6}
\end{eqnarray}

\noindent
Here $ t_c $ is the cut-off time ($\simeq 1/\omega$)
as has been mentioned earlier in the discussions preceding Eq.(\ref{eq60}).
Similarly, after averaging the last term of Eq.(\ref{eqa5}) becomes,

\begin{eqnarray}
[ {\rm last \; term \; of \; Eq.(\ref{eqa5}) } ]
& = & \sum_{n = -\infty}^\infty \sum_{l = 3}^\infty
\mu_n \exp ( in \phi ) \Upsilon_{1l} (0) t_c \tau \delta_{n,0}   
\nonumber  \\
& = & - \mu_0 t_c \tau f_0^\prime
\label{eqb7}
\end{eqnarray}

\noindent
where we have used Eq.(\ref{eq43}) and Eq.(\ref{eq59e}). 

Now consider the last integral of the right hand side of Eq.(\ref{eqa15}). 
It is of the form
$ \int_0^\tau ds \int_0^s ds_1 g (s) g (s_1) 
\exp ( in\omega s + il\omega s_1 ) $.
Using Eq.(\ref{eqb2}), it can be cast in the following form:- 

\begin{eqnarray}
& & \int_0^\tau ds \int_0^s ds_1 g (s) g (s_1) 
\exp ( in\omega s + il\omega s_1 ) 
\nonumber  \\
& = & \int_0^\tau ds g(s) \exp (in\omega s) 
\int_0^s ds_1  g (s_1) \exp (il\omega s_1)  
\nonumber  \\
& = & \int_0^\tau ds \left [ -\sum_{m = 0}^\infty \sum_{k = 3}^\infty 
s^m \Upsilon_{mk} (0) \right ] \exp (in\omega s) 
\int_0^s ds_1 \left [ -\sum_{m_1 = 0}^\infty \sum_{k_1 = 3}^\infty 
s_1^{m_1} \Upsilon_{m_1 k_1} (0) \right ] \exp (il\omega s_1)
\nonumber  \\
& = &  \sum_{m = 1}^\infty \sum_{k = 3}^\infty 
\sum_{m_1 = 1}^\infty \sum_{k_1 = 3}^\infty \Upsilon_{1k} (0)
\Upsilon_{1 k_1} (0) \int_0^\tau ds s^m \exp (in\omega s) 
\int_0^s ds_1 s_1^{m_1} \exp (il\omega s_1)
\nonumber   \\
& \simeq &  \sum_{k = 3}^\infty \sum_{k_1 = 3}^\infty \Upsilon_{1k} (0)
\Upsilon_{1 k_1} (0) \int_0^\tau ds s \exp (in\omega s) 
\int_0^s ds_1 s_1 \exp (il\omega s_1)  \; \; \; \; \;
[{\rm using \; Eq.(\ref{eqb5})}]
\nonumber   \\
& \simeq &  \sum_{k = 3}^\infty \sum_{k_1 = 3}^\infty \Upsilon_{1k} (0)
\Upsilon_{1 k_1} (0) \int_0^\tau ds s \exp (in\omega s) 
\left [ \frac { s \exp (il\omega s) }{ il\omega } \right ]
\label{eqb8}
\end{eqnarray}

\noindent
where evaluating the third bracket of the last expression we have neglected
terms of order $ O ( 1/ \omega^2 ) $. The last expression yields terms of
negligible contribution due to the 
reasons given at the onset of Eq.(\ref{eqb5}).
Integrals involving three $ g $ functions can be similarly shown to be
negligibly small.


\section{Treatment of integrands of $I_4$ to $I_7$ of Appendix-A}

In the discussions that appeared in between Eq.(\ref{eqa22g}) 
and Eq.(\ref{eqa23}) we mentioned of four
types of integrands which appear in the calculations involving the 
evaluations of the integrals $ I_4 $ to $ I_7 $ ( Eqs.(\ref{eqa22d})-
(\ref{eqa22g}) ). Let us recall their forms again. They are,

\begin{eqnarray*}
(1) \; \langle F(.) F(.) F(.) \rangle_S, \; \;
(2) \; \langle F(.) \rangle_S g(.) g(.), \; \;
(3) \; \langle F(.) F(.) \rangle_S g(.), \; \;
(4) \; g(.) g(.) g(.)
\end{eqnarray*}

\noindent
It was reasoned there that types $ (1), \; (2) $ and $ (4) $ yield terms
of negligible contributions (see Appendix-B). Here we clarify the way of
dealing with type $ (3) $. The specific forms of integrals involving this
type of integrands are:-

\begin{eqnarray*}
& (A) & \; \; \; \int_0^\tau ds \int_0^s ds_1 \int_0^{s_1} ds_2 
\langle F(s) F(s_1) \rangle_S g(s_2) \exp [ i\omega ( js + ns_1 + ls_2 )]  \\
& (B) & \; \; \; \int_0^\tau ds \int_0^s ds_1 \int_0^{s_1} ds_2 
\langle F(s) F(s_2) \rangle_S g(s_1) \exp [ i\omega ( js + ns_1 + ls_2 )]  
\; \; .
\end{eqnarray*}

\noindent
These two integrals occur in $ I_4 $.

\begin{eqnarray*}
(C) \; \; \; \int_0^\tau ds \int_0^s ds_1 \int_0^{s_1} ds_2 
\langle F(s) F(s_2) \rangle_S g(s_2) \exp [ i\omega ( js + ns_1 + ls_2 )]  
\; \; .
\end{eqnarray*}

\noindent
This, with type $ (A) $ above, occur in $ I_5 $. And lastly the integral,

\begin{eqnarray*}
(D) \; \; \; \int_0^\tau ds \int_0^s ds_1 \int_0^{s_1} ds_2 
\langle F(s_1) F(s_2) \rangle_S g(s) \exp [ i\omega ( js + ns_1 + ls_2 )]
\end{eqnarray*}

\noindent
occurs in $ I_6 $ and $ I_7 $.

All the integrals from $ (A) $ to $ (D) $ lead to negligible 
contributions. We establish this by showing the case of, say, type $ (B) $.
For doing this we invoke the quantum fluctuation-dissipation relation
from Eq.(\ref{fdr}) as,

\begin{eqnarray*}
\langle F(s)F(s_2) \rangle_S & = & \frac {1}{2} \int_0^{\infty} d\omega '
\kappa(\omega ') \rho(\omega ') \hbar \omega ' \left [ \coth \frac 
{\hbar \omega '}{2k_BT} \right ] \cos \omega ' ( s - s_2 ) \\
& = & \int_0^{\infty} d\omega '  G(\omega ') \cos \omega ' ( s - s_2 )  \\
{\rm with} \; \; \; \; \; \; \; \; \; \; & &  \\
G(\omega ') & = &
\frac {1}{2} \kappa(\omega ') \rho(\omega ') \hbar \omega '
\left [ \coth \frac {\hbar \omega '}{2k_BT} \right ] 
\end{eqnarray*}

\noindent
where, a superscript ``$'$'' has been added as a superscript to 
$ \omega $ to denote the frequency of the bath modes.

\begin{equation}
\label{eqc1}
\langle F(s)F(s_2) \rangle_S = \frac {1}{2} \int_0^{\infty} d\omega '  
G(\omega ') [ \exp \{ i\omega ' (s - s_2) \} + 
\exp \{ -i\omega ' (s - s_2) \} ]
\end{equation}

\noindent
When Eq.(\ref{eqc1}) is put in type $ (B) $ integral along with Eq.(\ref{eqb2}),
we get the value of the integral as,

\begin{equation}
\label{eqc2}
[ {\rm type \; B} ] = f_0^\prime \int_0^{\infty} d\omega ' G(\omega ')
\left [ \frac { 2\omega ' }{ n^2 \omega^2 - \omega'^2 } \right ]
t_c \tau \delta_{j+l, -n}
\end{equation}          

\noindent
This being of higher order in $ 1/ \omega $, can be discarded. 
It is easy to show that the other three
types of integrals also lead to negligible contributions. The procedure
is the same as that adopted above in the case of type $ (B) $.

\end{appendix}



\begin{thebibliography}{99}

\bibitem{hak} H.A. Kramers, 
Physica (Amsterdam) {\bf 7}, 284 (1940).

\bibitem{diau} E.W.-G. Diau, J.L. Herek, Z.H. Kim, and A.H. Zewail,
Science {\bf 279}, 847 (1998).

\bibitem{castelman} A.W. Castelman, D.E. Folmer, and E.S. Wisniewski,
Chem. Phys. Lett. {\bf 287}, 1 (1998).

\bibitem{htb} P. H\"anggi, P. Talkner, and M. Borkovec,
Rev. Mod. Phys. {\bf 62}, 251 (1990), and the references given therein.

\bibitem{vim} V.I. Mel'nikov,
Phys. Rep. {\bf 209}, 1 (1991).

\bibitem{fh} {\it Activated Barrier Crossing: Applications in Physics,
Chemistry, and Biology}, 
edited by G.R. Fleming and P. H\"anggi
(World Scientific, Singapore, 1993).

\bibitem{th1} {\it New Trends in Kramers' Reaction Rate Theory},
edited by P. Talkner and P. H\"anggi
(Kluwer, Dordrecht, 1995).

\bibitem{gh} R.F. Grote and J.T. Hynes,
J. Chem. Phys. {\bf 73}, 2715 (1980).

\bibitem{hm} P. H\"anggi and F. Mojtabai,
Phys. Rev. A {\bf 26}, 1168 (1982).

\bibitem{cn} B. Carmeli and A. Nitzan,
J. Chem. Phys. {\bf 79}, 393 (1983).

\bibitem{gt} R. Graham and T. T\'el,
Phys. Rev. Lett. {\bf 52}, 9 (1984);
Phys. Rev. A {\bf 31}, 1109 (1985).

\bibitem{stein} M. Array\'as, M.I. Dykman, R. Mannella, 
P.V.E. McClintock, and N.D. Stein,
Phys. Rev. Lett. {\bf 84}, 5470 (2000).

\bibitem{bcb} B.C. Bag and D.S. Ray,
Phys. Rev. E {\bf 61}, 3223 (2000); {\bf 62}, 4409 (2000).

\bibitem{jsl} J.S. Langer,
Ann. Phys. (N.Y.) {\bf 54}, 258 (1969).

\bibitem{bpz} A.M. Berezhkovskii, E. Pollak, and V.Yu. Zitserman,
J. Chem. Phys. {\bf 97}, 2422 (1992).

\bibitem{dg} C.R. Doering and J.C. Gadoua,
Phys. Rev. Lett. {\bf 69}, 2318 (1992).

\bibitem{nsp} R.N. Mantegna and B. Spagnolo,
Phys. Rev. Lett. {\bf 76}, 563 (1996);
V.N. Smelyanskiy, M.I. Dykman, and B. Golding,
{\it ibid}. {\bf 82}, 3193 (1999);
J. Lehmann, P. Reimann, and P. H\"anggi,
{\it ibid}. {\bf 84}, 1639 (2000); Phys. Rev. E {\bf 62}, 6282 (2000);
N. G. Stocks,
Phys. Rev. Lett. {\bf 84}, 2310 (2000);
R.S. Maier and D.L. Stein,
{\it ibid}. {\bf 86}, 3942 (2001).

\bibitem{tr} 
A. Ajdari and J. Prost,
C. R. Acad. Sci. Paris II {\bf 315}, 1635 (1992);
M.O. Magnasco,
Phys. Rev. Lett. {\bf 71}, 1477 (1993);
R.D. Astumian and M. Bier,
{\it ibid}. {\bf 72}, 1766 (1994);
R. Bartussek, P. H\"anggi, and J.G. Kissner,
Europhys. Lett. {\bf 28}, 459 (1994);
M.M. Millonas and M.I. Dykman,
Phys. Lett. A {\bf 183}, 65 (1994).

\bibitem{ken}
J.M.R. Parrondo and P. Espa\~nol,
Am. J. Phys. {\bf 64}, 1125 (1996);
K. Sekimoto,
J. Phys. Soc. Japan {\bf 66}, 1234 (1997).

\bibitem{mm} R.D. Astumian,
Science {\bf 276}, 917 (1997);
F. J\"ulicher, A. Ajdari, and J. Prost,
Rev. Mod. Phys. {\bf 69}, 1269 (1997);
N. Thomas and R.A. Thornhill,
J. Phys. D: Appl. Phys. {\bf 31}, 253 (1998);
P. Reimann,
Phys. Rep. {\bf 361}, 57 (2002).

\bibitem{garg} A.J. Leggett, S. Chakravarty, A.T. Dorsey, M.P.A. Fisher,
A. Garg, and W. Zwerger,
Rev. Mod. Phys. {\bf 59}, 1 (1987).

\bibitem{ingold} H. Grabert, P. Schramm, and G.L. Ingold,
Phys. Rep. {\bf 168}, 115 (1988).

\bibitem{uw} U. Weiss,
{\it Quantum Dissipative Systems} (World Scientific, Singapore, 1993).

\bibitem{jrc} J. Ray Chaudhuri, B.C. Bag, and D.S. Ray,
J. Chem. Phys. {\bf 111}, 10852 (1999).

\bibitem{kt} D.J. Tannor and D. Cohen,
J. Chem. Phys. {\bf 100}, 4932 (1994);
D. Kohen and D.J. Tannor, 
{\it ibid}. {\bf 103}, 6013 (1995).

\bibitem{srl} J.M. Sancho, A.H. Romero, and K. Lindenberg,
J. Chem. Phys. {\bf 109}, 9888 (1998);
K. Lindenberg, A.H. Romero, and J.M. Sancho,
Physica D {\bf 133}, 348 (1999).

\bibitem{mk} R. Metzler and J. Klafter,
Chem. Phys. Lett. {\bf 321}, 238 (2000);
Phys. Rep. {\bf 339}, 1 (2000), and the references given therein.

\bibitem{rsl} A.H. Romero, J.M. Sancho, and K. Lindenberg (unpublished),
Preprint cond-mat/0204389.

\bibitem{skb} S.K. Banik, J. Ray Chaudhuri, and D.S. Ray,
J. Chem. Phys. {\bf 112}, 8330 (2000).

\bibitem{endb} J. Ray Chaudhuri, S.K. Banik, B.C. Bag, and D.S. Ray,
Phys. Rev. E {\bf 63}, 061111 (2001).

\bibitem{bicout} D.J. Bicout, A.M. Berezhkovskii, A. Szabo, and
G.H. Weiss,
Phys. Rev. Lett. {\bf 83}, 1279 (1999);
D.J. Bicout, A.M. Berezhkovskii, and A. Szabo,
J. Chem. Phys. {\bf 114}, 2293 (2001).

\bibitem{dbbr} D. Banerjee, B.C. Bag, S.K. Banik, and D.S. Ray,
Phys. Rev. E {\bf 65}, 021109 (2002).

\bibitem{epw} E. Wigner,
Phys. Rev. {\bf 40}, 749 (1932);
M. Hillery, R.F. O'Connell, M.O. Scully, and E.P. Wigner,
Phys. Rep. {\bf 106}, 121 (1984).

\bibitem{aoc} A.O. Caldeira and A.J. Leggett,
Phys. Rev. Lett. {\bf 46}, 211 (1981);
Ann. Phys. (N.Y.) {\bf 149}, 374 (1983);
Physica A {\bf 121}, 587 (1983).

\bibitem{hw} P. H\"anggi and U. Weiss,
Phys. Rev. A {\bf 29}, 2265 (1984).

\bibitem{gh1} R.F. Grote and J.T. Hynes,
J. Chem. Phys. {\bf 77}, 3736 (1982).

\bibitem{th} T. Hesselroth,
Phys. Rev. E {\bf 48}, 46 (1993).

\bibitem{vm} V.I. Mel'nikov,
Zh. Eksp. Teor. Fiz. {\bf 87}, 663 (1984)
[Sov. Phys. JETP {\bf 60}, 380 (1984)];
Physica A {\bf 130}, 606 (1985).

\bibitem{lo} A.I. Larkin and Yu.N. Ovchinnikov,
J. Stat. Phys. {\bf 41}, 425 (1985).

\bibitem{rj} I. Rips and J. Jortner,
Phys. Rev. B {\bf 34}, 233 (1986).

\bibitem{chow} K.S. Chow and V. Ambegaokar,
Phys. Rev. B {\bf 38}, 11168 (1988).

\bibitem{dekker} H. Dekker,
Phys. Rev. A {\bf 38}, 6351 (1988);
H. Dekker and A. Maassen van den Brink,
Phys. Rev. E {\bf 49}, 2559 (1994).

\bibitem{griff} U. Griff, H. Grabert, P. H\"anggi, and P.S. Riseborough,
Phys. Rev. B {\bf 40}, 7295 (1989).

\bibitem{whl} W.H. Louisell,
{\it Quantum Statistical Properties of Radiation}
(Wiley, New York, 1973)

\bibitem{bbr} S.K. Banik, B.C. Bag, and D.S. Ray,
Phys. Rev. E {\bf 65}, 051106 (2002).

\bibitem{rv} H. Risken and K. Vogel, in
{\it Far From Equilibrium Phase Transition}, edited by L. Garrido,
Lecture Notes in Physics, Vol.~{319} (Springer-Verlag, Berlin, 1988).

\bibitem{lax} M. Lax,
Rev. Mod. Phys. {\bf 38}, 541 (1966).

\bibitem{bhl} M. B\"uttiker, E.P. Harris, and R. Landauer,
Phys. Rev. B {\bf 28}, 1268 (1983).

\bibitem{troe} H. Hippler, K. Luther, and J. Troe,
Ber. Bunsenges. Phys. Chem. {\bf 77}, 1104 (1973);
K. Luther, J. Schroeder, J. Troe, and U. Unterberg,
J. Phys. Chem. {\bf 84}, 3072 (1980);
B. Otto, J. Schroeder, and J. Troe,
J. Chem. Phys. {\bf 81}, 202 (1984).

\bibitem{hej} D.L. Hasha, T. Eguchi, and J. Jonas,
J. Am. Chem. Soc. {\bf 104}, 2290 (1982).

\bibitem{jt} J. Troe,
J. Phys. Chem. {\bf 90}, 357 (1986);
J. Schroeder and J. Troe,
Annu. Rev. Phys. Chem. {\bf 38}, 163 (1987).

\bibitem{hara} K. Hara, N. Ito, and O. Kajimoto,
J. Chem. Phys. {\bf 110}, 1662 (1999).

\bibitem{cmc} A.N. Cleland, J.M. Martinis, and J. Clarke,
Phys. Rev. B {\bf 37}, 5950 (1988).

\bibitem{rz} R. Zwanzig,
J. Stat. Phys. {\bf 9}, 215 (1973).

\bibitem{west} K. Lindenberg and B.J. West,
{\it The Nonequilibrium Statistical Mechanics of Open and Closed Systems}
(VCH, New York, 1990).

\bibitem{jkb} J.K. Bhattacharjee,
{\it Statistical Physics: Equilibrium and Non-Equilibrium Aspects}
(Allied Publishers, New Delhi, 1997).

\bibitem{akp} A.K. Pattanayak and W.C. Schieve,
Phys. Rev. E {\bf 50}, 3601 (1994).

\bibitem{sm} B. Sundaram and P.W. Milonni,
Phys. Rev. E {\bf 51}, 1971 (1995).

\bibitem{risken} H. Risken,
{\it The Fokker-Planck Equation}
(Springer-Verlag, Berlin, 1989)

\bibitem{fv} R.P. Feynman and F.L. Vernon,
Ann. Phys. (N.Y.) {\bf 24}, 118 (1963).

\bibitem{hibbs} R.P. Feynman and A.R. Hibbs,
{\it Quantum Mechanics and Path Integrals}
(McGraw-Hill, New York, 1965).

\bibitem{topaler} M. Topaler and N. Makri,
J. Chem. Phys. {\bf 101}, 7500 (1994), and the references given therein.

\bibitem{liao} J.-L. Liao and E. Pollak,
J. Chem. Phys. {\bf 116}, 2718 (2002).

\bibitem{Creswick} R. Creswick,
Mod. Phys. Lett. B {\bf 9}, 693 (1995).


\end{thebibliography}
\end{document}